\documentclass[11pt]{article}

\usepackage[body={16cm,23cm}]{geometry}
\usepackage{cite}
\usepackage{amsmath,amssymb}
\usepackage{amsfonts}
\usepackage[mathscr]{eucal}
\usepackage{epic}
\usepackage{eepic}
\usepackage{enumerate}
\usepackage{graphicx}
\usepackage[hang,small,bf
]{caption}
\newcounter{app}
\newcounter{sapp}[app]
\def\theapp{\Alph{app}}
\newcommand{\app}[1]{
\refstepcounter{app}{\vspace{7mm}
\noindent\Large\bf Appendix
\theapp.
 \ #1 \par \vspace{5mm}}
\setcounter{equation}{0}
\def\theequation{\Alph{app}.\arabic{equation}}}

\newcommand{\ds}{\displaystyle}
\def\EXP{\textrm{{\large e}}}

\newcommand{\ii}{\mathsf{i}}

\renewcommand{\t}{\theta}
\newcommand{\crs}{\eta}
\newcommand{\norma}{F}
\newcommand{\Gfun}{\Phi}
\newcommand{\be}{\stackrel{\textrm{def}}{=}}
\renewcommand{\Im}{\mathrm{Im}}
\renewcommand{\Re}{\mathrm{Re}}
\font\cyr=wncyr10 
\newcommand{\lobachevski}{\operatorname{\mbox{\cyr L}}}

\renewcommand{\author}[1]{\large\rm #1\\ \bigskip}
\newcommand{\address}[1]{{\normalsize\it #1\\}\bigskip}
\renewcommand{\title}[1]{\bigskip\bigskip\Large\bf #1\bigskip\bigskip\\}

\begin{document}

\vglue 2cm

\begin{center}

\title{
Faddeev-Volkov solution of the Yang-Baxter Equation\\
and Discrete Conformal Symmetry%
}

\vspace{1cm}

\author{        Vladimir V. Bazhanov\footnote[1]{email:
                {\tt Vladimir.Bazhanov@anu.edu.au}},
                Vladimir V. Mangazeev\footnote[2]{email:
                {\tt Vladimir.Mangazeev@anu.edu.au}},
                Sergey M. Sergeev\footnote[3]{email:
                {\tt Sergey.Sergeev@anu.edu.au}}}

\address{Department of Theoretical Physics,\\
         Research School of Physical Sciences and Engineering,\\
    Australian National University, Canberra, ACT 0200, Australia.}

\end{center}






\vspace{1cm}

\begin{abstract}
The Faddeev-Volkov solution of the star-triangle relation is
connected with the modular double of the quantum group
$U_q(sl_2)$. It defines an Ising-type lattice model with positive
Boltzmann weights  where the spin variables take continuous values
on the real line. The free energy of the model is exactly
calculated in the thermodynamic limit. The model describes
quantum fluctuations of circle patterns and the associated
discrete conformal transformations connected
with the Thurston's discrete analogue of the Riemann mappings
theorem.
In
particular, in the quasi-classical limit the model precisely
describe the geometry of integrable circle patterns
with prescribed intersection angles.
\end{abstract}

\newpage
\section{Introduction}
The Faddeev-Volkov solution \cite{Volkov:1992,FV:1993,Faddeev:1994}
of the Yang-Baxter equation \cite{YBEbook}
possesses many remarkable properties. From algebraic point of view it
is distinguished by its {\em modular duality}.
The solution is related with non-compact representations of the
modular double \cite{Faddeev:1999} of the quantum group
$U_q({sl}_2)\otimes U_{\tilde{q}}({sl}_2)$, where
$q=e^{i\pi b^2}$ and $\tilde{q}=e^{-i\pi/ b^2}$.
It was studied in connection with the lattice Liouville and
sinh-Gorgon models \cite{FKV:2001, Kharchev:2002, Volkov:2005,
Bytsko:2006}. Note, that the parameter
$b$ above is related to the Liouville central charge
$c_L=1+6\,(b+b^{-1})^2$.

It is well known that every solution of the Yang-Baxter equation
defines an integrable model of statistical mechanics on a
two-dimensional lattice.
The Faddeev-Volkov model, defined by their solution,
has strictly positive Boltzmann weights.
It is an Ising-type model with continuous spin variables $\sigma_i\in
{\mathbb R}$, taking
arbitrary real values.
Its partition function is given by the integral
\begin{equation}
Z=\int \EXP^{\displaystyle-{\mathscr E}[\sigma]}\ \prod_i d \sigma_i,
\label{ZFV}
\end{equation}
where the energy ${\mathscr E}[\rho]$ is a convex and bounded
from below function of the spin variables
$\sigma=\{\sigma_1,\sigma_2, \ldots\}$.
In this paper we study this model and
reveal  its remarkable
connections to discrete geometry,
namely,
to {\em circle patterns} and
{\em discrete conformal transformations}
(for an introduction see a popular review \cite{St1} and a monograph
\cite{Steph:2005}).
Recently these topics attracted much attention due to
Thurston's circle packing approach to the discrete
Riemann mapping theorem
\cite{T1, RS, HS, S} and the discrete uniformization theorem
\cite{T2, MR, BeaS, He}.

In this approach {\em circle packings} and, more generally, {\em
  circle patterns} (see Sect.~5 below) serve as  discrete analogues of
  conformal transformations, where the circle radii  can be thought as
  the local dilatation factors. They satisfy certain integrable non-linear
difference equations which play the
role of discrete Cauchy-Riemann conditions.
These equations are called the {\em cross-ratio equations}
\cite{BP1,NC,BoSur}.  In the
simplest case they reduce to the Hirota form of the discrete sine-Gordon
equation \cite{hirota3,FV:1994}. (Curiously enough, the continuous
  Riemann mapping problem for domains with a piece-wise smooth boundary
  is also connected to the Hirota equations \cite{Wiegmann}, but of
  the different type
  \cite{Hirota2}.) An appropriate variational approach for the circle
  patterns was
developed by Bobenko and Springborn \cite{BSp}. Their
action ${\mathscr A}[\rho]$ is a positive definite
function of the logarithmic radii of the circles
$\rho=\{\rho_1,\rho_2\ldots\}$, \ $\rho_i=\log r_i$.
We show that the leading quasi-classical asymptotics of
the energy functional in \eqref{ZFV}
\begin{equation}
{\mathscr E}[\sigma]
\simeq {\frac{1}{2\pi b^2}\, {\mathscr A}[2\pi b \sigma]},
\qquad b\to0,\qquad c_L\to+\infty,
\end{equation}
precisely coincide with this action, provided logarithmic radii of the circles
are identified with the (rescaled) spin variables in \eqref{ZFV},
namely $\rho_i=2\pi b \sigma_i$.
Thus, at $c_L\to+\infty$, stationary points of the integral
\eqref{ZFV} correspond to circle patterns. At finite values of $c_L$
the model describes quantum fluctuations of these patterns
and the associated discrete conformal transformation. Given that
the circle radii define local dilatation factors, the Faddeev-Volkov model
describes a {\em quantum discrete dilaton}.

The continuous quantum field theory with the conformal symmetry
\cite{BPZ84} has remarkable applications in physics and mathematics
\cite{ISZ88}.  It would be interesting to understand which aspects of
the continuous conformal  field theory can be transferred to the
discrete case.  

The organization of the paper is as follows. In Sect.~2
we present definitions and discuss properties of the Faddeev-Volkov model.
In Sect.~3 we consider $Z$-invariant lattices and rhombic tilings
which are closely related to the combinatorics of the circle patterns. The
quasi-classical limit is considered in Sect.~4. An introduction to
circle patterns is given in Sect.~5. We review the derivation of the
cross-ratio equations and show that they are exactly coincide with the
quasi-classical  equation of motion  in  the Faddeev-Volkov model.
Some open problem are discussed in the Conclusion.
In Appendix~A we discuss some implications of the star-triangle equation
and $Z$-invariance to the circle patterns and volumes of polyhedra in
the Lobachevskii 3-space.
The Appendix~B contains useful information on the special functions
used in this paper.
\section{The Faddeev-Volkov model}
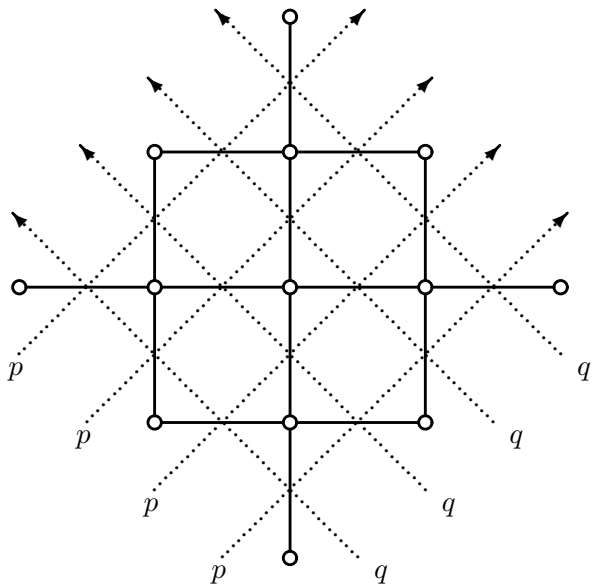
\begin{figure}[thb]
\begin{center}
\setlength{\unitlength}{0.09mm}
\begin{picture}(1000,1000)
\Thicklines
\dottedline[$\cdot$]{12}(600,100)(100,600)\put(100,600){\vector(-1,1){10}}
\put(610,70){ $q$}
\dottedline[$\cdot$]{12}(700,200)(200,700)\put(200,700){\vector(-1,1){10}}
\put(710,170){ $q$}
\dottedline[$\cdot$]{12}(800,300)(300,800)\put(300,800){\vector(-1,1){10}}
\put(810,270){ $q$}
\dottedline[$\cdot$]{12}(900,400)(400,900)\put(400,900){\vector(-1,1){10}}
\put(910,370){ $q$}
\dottedline[$\cdot$]{12}(100,400)(600,900)\put(600,900){\vector(1,1){10}}
\put(70,370){ $p$}
\dottedline[$\cdot$]{12}(200,300)(700,800)\put(700,800){\vector(1,1){10}}
\put(170,270){ $p$}
\dottedline[$\cdot$]{12}(300,200)(800,700)\put(800,700){\vector(1,1){10}}
\put(270,170){ $p$}
\dottedline[$\cdot$]{12}(400,100)(900,600)\put(900,600){\vector(1,1){10}}
\put(370,70){ $p$}
\put(500,900){\circle{20}}
\put(300,700){\circle{20}}\put(500,700){\circle{20}}\put(700,700){\circle{20}}
\put(100,500){\circle{20}}\put(300,500){\circle{20}}\put(500,500){\circle{20}}
\put(700,500){\circle{20}}\put(900,500){\circle{20}}
\put(300,300){\circle{20}}\put(500,300){\circle{20}}\put(700,300){\circle{20}}
\put(500,100){\circle{20}}
\allinethickness{.4mm}
\path(500,110)(500,290)\path(500,310)(500,490)\path(500,510)(500,690)\path(500,710)(500,890)
\path(300,310)(300,490)\path(300,510)(300,690)
\path(700,310)(700,490)\path(700,510)(700,690)

\path(110,500)(290,500)\path(310,500)(490,500)\path(510,500)(690,500)\path(710,500)(890,500)
\path(310,300)(490,300)\path(510,300)(690,300)
\path(310,700)(490,700)\path(510,700)(690,700)
\end{picture}
\end{center}
\caption{The square lattice (solid lines) and its medial lattice
  (dashed lines).}\label{fig0}
\end{figure}

\begin{figure}[hbt]
\begin{center}
\setlength{\unitlength}{.17in} \thicklines
\def\punit#1{\hspace{#1\unitlength}}
\def\pvunit#1{\vspace{#1\unitlength}}
\begin{picture}(4,9)\put(0,1)
{\begin{picture}(4,8)(0,-4.0)
\put(-6.5,0){\line(1,0){5.1}} \put(-3.4,0){\line(1,0){1.0}}
\put(-6.5,-3.2){\makebox(0,0)[b]{\small \mbox{$p$}}}
\put(-1.5,-3.2){\makebox(0,0)[b]{\small \mbox{$q$}}}
\put(-0.4,-0.2){\makebox(0,0)[b]{\small \mbox{$b$}}}
\put(-7.6,-0.2){\makebox(0,0)[b]{\small \mbox{$a$}}}
\put(-4.2,-5.6){\makebox(0,0)[b]{\small \mbox{$W\! _{\! p-q}
(a-b)$}}} \multiput(-6.4,-2.2)(0.2,0.2){22}{\makebox(0,0)[b]{
\mbox{${.}$}}} \multiput(-6.4,2.0)(0.2,-0.2){22}{\makebox(0,0)[b]{
\mbox{${.}$}}} \put(-6.8,0){\circle{.5}} \put(-1.2,0){\circle{.5}}
\put(8.0,-2.4){\line(0,1){4.6}} \put(8.0,0.3){\line(0,1){1.0}}
\put(5.5,-3.0){\makebox(0,0)[b]{\small \mbox{$p$}}}
\put(10.5,-3.0){\makebox(0,0)[b]{\small \mbox{$q$}}}
\put(8.0,3.0){\makebox(0,0)[b]{\small \mbox{$b$}}}
\put(7.4,-3.6){\makebox(0,0)[b]{\small \mbox{$a$}}}
\put(7.8,-5.6){\makebox(0,0)[b]{\small \mbox{$\overline{W}\! _{\!
p-q} (a-b)$}}}
\multiput(5.8,-2.2)(0.2,0.2){22}{\makebox(0,0)[b]{\mbox{${.}$}}}
\multiput(5.8,2.2)(0.2,-0.2){23}{\makebox(0,0)[b]{\mbox{${.}$}}}
\put(8.0,-2.6){\circle{.5}} \put(8.0,2.4){\circle{.5}}
\put(-1.8,2.2){\vector(1,1){0.3}}
\put(-6.3,2.2){\vector(-1,1){0.3}}
\put(10.2,2.2){\vector(1,1){0.3}}
\put(5.9,2.2){\vector(-1,1){0.3}}
\end{picture}}\end{picture}\end{center}
\caption{Typical horizontal and vertical edges and the corresponding
  Boltzmann weights.}\label{fig1}
\end{figure}
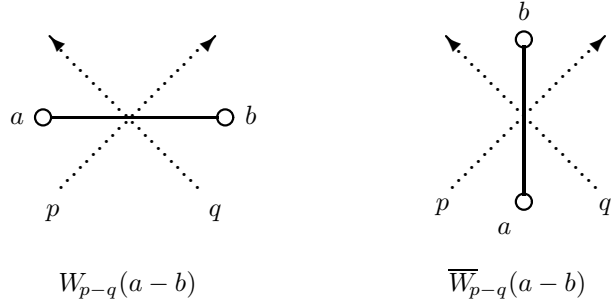

Consider an Ising-type model on the square lattice, shown in
Fig.\ref{fig0}. Each site $i$ of the lattice is assigned with a spin
variable $\sigma_i\in {\mathbb R}$,
taking continuous real values. Two spins $a$ and $b$
interact only if they are
adjacent (connected with an edge of the lattice).
Typical
horizontal and vertical edges are shown in Fig.\ref{fig1}.
The corresponding Boltzmann weights are parametrized through the
so-called ``rapidity variables'' $p$ and $q$ associated with the
oriented dashed lines in Fig.\ref{fig0}. In our case the weights
depends only on the spin and rapidity differences $a-b$ and $p-q$,
where $a$ and $b$ are the spins at the ends of the edge.
We will denote them as $W_{p-q}(a-b)$ and $\overline{W}_{p-q}(a-b)$
for the horizontal and vertical edges, respectively.

The partition function is defined as
\begin{equation}
Z=\int  \prod_{(i,j)}
W_{p-q}(\sigma_i-\sigma_j)\ \prod_{(k,l)}
\overline{W}_{p-q}(\sigma_k-\sigma_l)\ \prod_i d\sigma_i\;.\label{Z-def}
\end{equation}
where the first product is over all horizontal edges $(i, j)$, the
second is over all vertical edges $(k,l)$. The integral is taken over
all the interior spins; the boundary spins are kept fixed.

Explicit expressions for the Boltzmann weights contain
two special functions. The first one,
\begin{equation}
\varphi(z)\;\be\;\exp\left(\ds \frac{1}{4}\int_{\mathbb{R}+\ii
0} \frac{\EXP^{-2\ii zw}}{\textrm{sinh}(wb)\textrm{sinh}(w/b)}\
\frac{dw}{w}\right)\;,\label{fi-def}
\end{equation}
is the Barnes double-sine function \cite{Barnes:1901}.
It was introduced to this field in \cite{Faddeev:1994}
as the non-compact version of the quantum dilogarithm
\cite{Faddeev:1993rs}. The second function has no special name,
\begin{equation}
\Gfun(z)\; \be \;\exp\left(\frac{1}{8}\int_{\mathbb{R}+\ii 0}
\frac{\EXP^{-2\ii z w}}
{\sinh(wb)\sinh(wb^{-1})\cosh(w(b+b^{-1}))}\
\frac{dw}{w}\right)\;,\label{Fi-def}
\end{equation}
but previously appeared \cite{FLZ} in the context of the quantum sinh-Gordon
model. The properties of these functions are briefly summarized in the
Appendix~\ref{phi-prop}.

The Boltzmann weights $W_{\theta\,}(s)$ and
$\overline{W}_{\t\,}(s)$ are defined as
\begin{equation}\label{W}
W_{\t\,}(s)\; \be \;\frac{1}{\norma_\t}\ \EXP^{2\crs \t s}\
\frac{\varphi(s+\ii \crs \t/\pi)}{\varphi(s-\ii \crs \t/\pi)}\;,\qquad\qquad
\overline{W}_{\t\,}(s)\; \be \; W_{\pi-\t}(s)\;,
\end{equation}
where $\t$ and $s$ stand for the rapidity and spin differences,
 respectively, and
the parameter $\crs$ reads
\begin{equation}\label{eta}
\crs=(b+b^{-1})/2\ .
\end{equation}
The normalization factor $\norma_\t$ has the form
\begin{equation}\label{F}
\norma_\t\; \be \;\EXP^{\ii\crs^2 \t^2/\pi + \ii \pi (1-8\crs^2)/24}\
\Gfun(2\ii \crs\/ \t/\pi)\;.
\end{equation}
The positivity requirement for the Boltzmann weights (point (b) below)
singles out  two different regimes
\begin{equation}
{\rm (i)}\ b>0,\qquad
{\rm or \ \ \ \ \ \ (ii)}\  |b|=1, \qquad \Im(b^2)>0.\label{regimes}
\end{equation}
The following consideration apply to either of these cases.
Note that due to the symmetry $b\leftrightarrow b^{-1}$, for the regime
(i) it is enough to consider the range $0<b\le 1$.
The weights  (\ref{W}) possess the following key properties:
\begin{enumerate}[(a)]
\item {\sl Symmetry:} $W_\t(s)$ and $\overline{W}_\t(s)$
are even functions of $s$. The definition \eqref{W} implies
\begin{equation}
\overline{W}_{\t\,}(s)= W_{\pi-\t}(s)\;.\label{crossing-sym}
\end{equation}
\item {\sl Positivity:} when the parameter $b$ belongs to either of
  the regimes \eqref{regimes}
the functions
$W_{\t\,}(s)$ and $\overline{W}_\t(s)$
are real and positive for $0<\t<\pi$ and real $s$ .
They have no zeros and poles on the real axis of $s$.
\item {\sl Asymptotics:}
\begin{equation}
W_{\t\,}(s)\; \simeq \; \norma_\t^{-1}\EXP^{-2\crs
\t |s|}\Big(1+O(\EXP^{-2\pi \beta |s|})\Big),\qquad {s\to \pm \infty}
 \;.\label{decay}
\end{equation}
where $\beta=\min(\mbox{Re}\, b, \mbox{Re}\, b^{-1})$.
\item {\sl Star-triangle relation:}
\begin{equation}\label{str}
\int_\mathbb{R} d\sigma \ \overline{W}_{ q-r\,}(a-\sigma) \ W_{
p-r\,}(c-\sigma) \ \overline{W}_{ p-q\, } (\sigma-b) \ =\ W_{
p-q\,}(c-a) \overline{W}_{ p-r\,}(a-b) \ W_{q-r\,}(c-b)\;.
\end{equation}
\item {\sl Inversion relations:}
\begin{equation}\label{inv}
W_{\t\,}(a-b) W_{-\t\,}(a-b)\;=\;1\;,\qquad \lim_{\varepsilon\to
0^+}\int_{\mathbb{R}} \; dc\; \overline{W}_{\ii
t+\varepsilon\,}(a-c) \overline{W}_{-\ii t
+\varepsilon\,}(c-b)\;=\;\delta(a-b)\;,
\end{equation}
where $t$ is real.
\item
{\sl Self-duality:}
\begin{equation}\label{Wp}
\overline{W}_{\t\,}(s)\;=\;\int_\mathbb{R}\  dx\ \EXP^{2\pi\ii xs}
\ W_{\t\,}(x)\;.
\end{equation}
\item
{\sl Initial conditions:}
\begin{equation}
W_\t(s)=1+O(\t) \quad\textrm{and}\quad \overline{W}_\t(s)\;=\;
\delta(s)\;+\; O(\t) \quad \textrm{as} \quad \t\to 0\;,
\end{equation}
where $\delta(s)$ is the $\delta$-function.
\end{enumerate}

The normalization factor $\norma_\t$ is the ``minimal'' solution of
the inversion and crossing symmetry relations,
\cite{Str79,Zam79,Bax82inv}
\begin{equation}\label{F-eq}
\norma_\t \norma_{-\t} \;=\; 1\;,\quad
\frac{\norma_\t}{\norma_{\pi-\t}}\;=\; \EXP^{-2\ii \crs^2 \t(\pi-\t)/\pi +
\ii\pi(1+4\crs^2)/12}\ \varphi\left(\frac{2\ii \crs \t}{\pi} -\ii
\eta\right)\;,
\end{equation}
in the sense that it does not have poles and zeroes in the strip
$-\pi<{\rm Re} \,\t<\pi$; it is real and positive for real
$-\pi<\t<\pi$, and has a simple pole at $\t=\pi$ and a simple
zero at $\t=-\pi$. Note also that $\norma_0=1$. Equation
(\ref{Wp}) is a consequence of (\ref{rstr}) and the second relation
in (\ref{F-eq}).

Let $N$ be the total number of edges on the lattice. In the thermodynamic
limit, when $N$ is large, the partition function
grows as
\begin{equation}
\log Z= -f_{bulk} N -f_{boundary} \sqrt{N} + O(1), \qquad N\to \infty,
\end{equation}
where the first and second terms describe the bulk and boundary
contributions to the free energy. 
The inversion relation
arguments of \cite{Str79,Zam79,Bax82inv}
imply that {\em with our normalization of
Boltzmann weights} \eqref{W} the bulk free energy for the system \eqref{Z-def}
vanishes exactly
\begin{equation}
-f_{bulk}=\lim_{N\to\infty}\frac{1}{N}\log Z =0 \label{fzero}
\end{equation}
Thus, $\log Z$ can
only grow linearly in the number of boundary edges, which is
$O(\sqrt{N})$. Here we assume that the boundary spins are kept finite 
in the limit $N\to\infty$, otherwise \eqref{fzero} will not hold. 

The weights \eqref{W} attain their maxima at $s=0$.
In the quasi-classical domain $0<b\le 1$ the constant
$W_{\frac{\pi}{2}}(0)$ slowly varies between the values
\begin{equation}
W_{\frac{\pi}{2}}(0)\Big|_{b=0}=\EXP^{\frac{G}{\pi}}=1.3385\ldots,\qquad
W_{\frac{\pi}{2}}(0)\Big|_{b=1}=\sqrt{2}=1.4142\ldots,
\end{equation}
where $G=0.915965\ldots$ is the
Catalan's constant. When the parameter $b$ on the unit
circle this constant
monotonically increases as $\arg(b)$ varies from $0$ to $\pi/2$ and
diverges when $b\to i$,
\begin{equation}
W_{\frac{\pi}{2}}(0)\simeq
(2\pi \crs)^{-1/2}\,\frac{\Gamma(\frac{1}{4})}
{\Gamma(\frac{3}{4})}+O\left(\crs^{1/2}\right), \qquad b\to i\ .
\end{equation}
Other properties of the Faddeev-Volkov model are considered in a
separate publication \cite{BMS07}.

\section{Z-invariant lattices and rhombic tilings}
The above considerations were explicitly carried out
for the homogeneous square lattice (shown in Fig.1).
However, having in mind connections with the discrete conformal
transformations considered below, it is useful to reformulate
the results for more general ``Z-invariant''
lattices  \cite{Bax1}. A brief review of this important
notion with an account of some subsequent developments is given below.

\begin{figure}[htb]
\begin{center}
\includegraphics[scale=0.5]{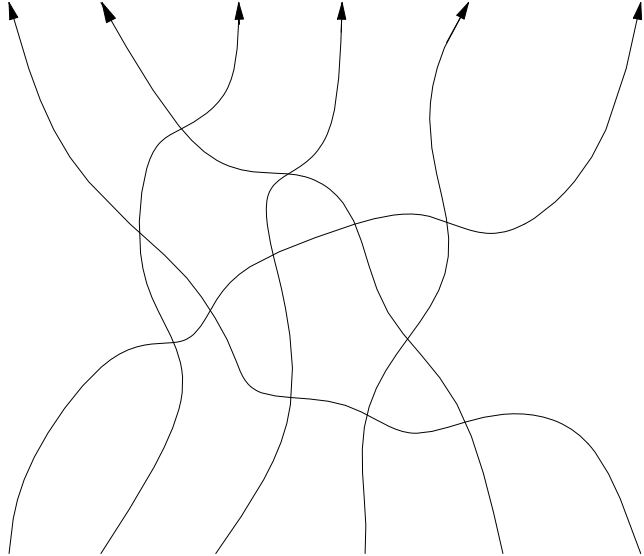}
\caption{A set of directed lines going from the bottom to the top
of the figure.} \label{fig-net}
\end{center}
\end{figure}

Following \cite{Bax2} consider a finite set of $L$ directed lines
forming a graph $\mathscr{L}$ of the type shown in Fig.~\ref{fig-net}.
The lines (in this case six) head generally from
the bottom of the graph to the top, intersecting one another  on
the way. They can go locally downwards, but there can be no
closed directed paths in $\mathscr{L}$. This means that one can
always distort $\mathscr{L}$, without changing its topology, so
that the lines always head upwards.
\begin{figure}[ht]
\begin{center}
\includegraphics[scale=0.6]{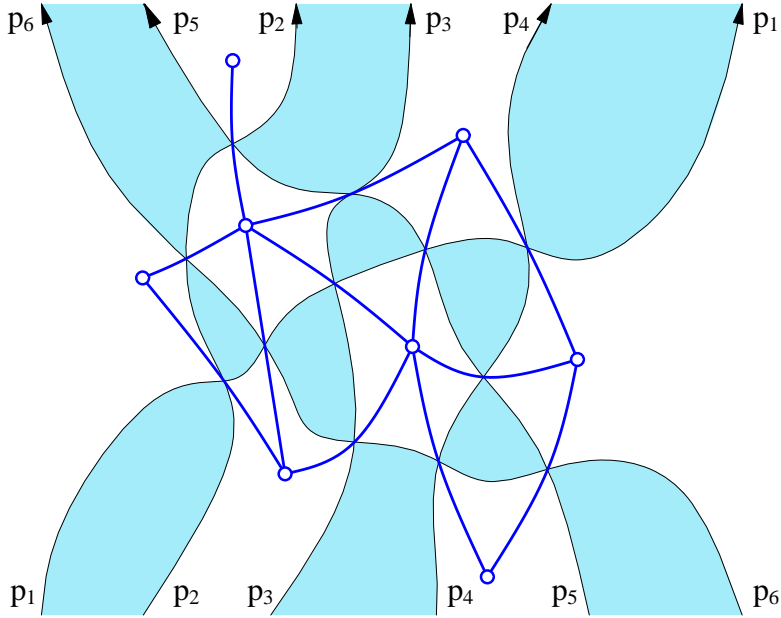}
\end{center}
\caption{The graph $\mathscr{G}$ formed by shading alternate faced
of Fig. \ref{fig-net} and putting sites on the unshaded faces.}
\label{fig-net2}
\end{figure}
Now shade alternative faces of $\mathscr{L}$ as in
Fig.~\ref{fig-net2}.
Form another graph $\mathscr{G}$ by placing a site
in each unshaded face, with edges connecting sites belonging to faces
that touch at a
corner. These sites and edges of $\mathscr{G}$ are represented in
Fig.~\ref{fig-net2} by open circles and solid lines,
respectively. For each intersection of lines in $\mathscr{L}$,
there is an edge of $\mathscr{G}$ passing through it, and
conversely. The graph $\mathscr{L}$ is the {\em medial} graph of $\mathscr{G}$.
Introduce additional notation. Let $F({\mathscr G})$, $E({\mathscr G})$
and $V({\mathscr G})$ denote the
sets of faces, edges and sites (vertices) of ${\mathscr G}$ and
$V_{int}({\mathscr G})$ the set of interior
sites of ${\mathscr G}$. The latter correspond to
interior faces of $\mathscr{L}$ (with a closed boundary).

Now we define a statistical mechanical spin model on $\mathscr{G}$.
With each line $\ell$ of ${\mathscr L}$
associate its own ``rapidity'' variable $p_{\ell}$.
At each site $i$ of $\mathscr{G}$ place a spin $\sigma_i\in\mathbb{R}$.
The edges of $\mathscr{G}$ are either of the first type in
Fig.~\ref{fig1},  or the second, depending on the arrangement
the directed rapidity lines with respect to the edge.
For each edge introduce a ``rapidity
difference variable'' $\theta_e$ defined as
\begin{equation}
\theta_e=\left\{\begin{array}{ll}
p-q,\ \ \ \ \ \ \ \mbox{for a first type edge,}\\
\\
\pi-p+q,\ \mbox{for a second type edge,}
\end{array}\right.\label{difvar}
\end{equation}
where $p$ and $q$ are the associated rapidities,
arranged as in Fig.~\ref{fig1}.
Each edge is assigned with
the Boltzmann weight factor $W_{\theta_e}(a-b)$, where $a$, $b$ are
the spins at the ends of the edge.

The partition is defined as an integral over all configurations
of the interior spins with the weight equal to the product
of the edge weights over all edges of ${\mathscr G}$,
\begin{equation}
Z=\int  \prod_{(i,j)\in E({\mathscr G})}
W_{\theta_{(i,j)}}(\sigma_i-\sigma_j)\  \prod_{i\in
V_{int}({\mathscr G})} d\sigma_i\;.\label{Z-def1}
\end{equation}
As before, the exterior spins are kept fixed.
Taking into account the symmetry \eqref{crossing-sym} one can easily see
that the expression (\ref{Z-def}) is just a particular case of
\eqref{Z-def1} for the homogeneous square lattice of Fig.~\ref{fig0}.

\begin{figure}[hbt]
\begin{center}
\includegraphics[scale=0.6]{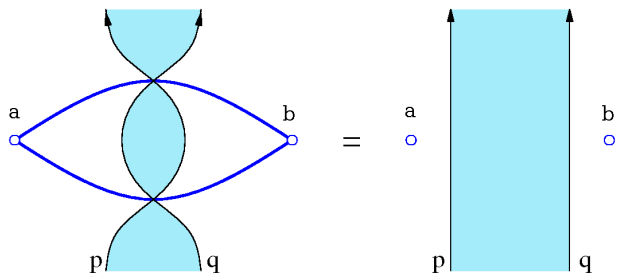}
\end{center}
\caption{A pictorial representation of the first inversion relation in
  \eqref{inv}.}
\label{fig-i1}
\end{figure}
\begin{figure}[hbt]
\begin{center}
\includegraphics[scale=0.5]{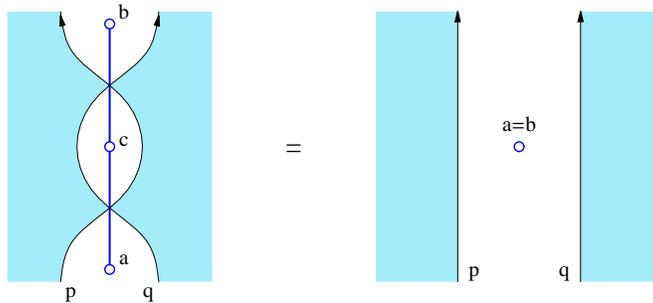}
\end{center}
\caption{A pictorial representation of the second inversion relation
  in \eqref{inv}.}
\label{fig-i2}
\end{figure}
\begin{figure}[hbt]
\begin{center}
\includegraphics[scale=0.6]{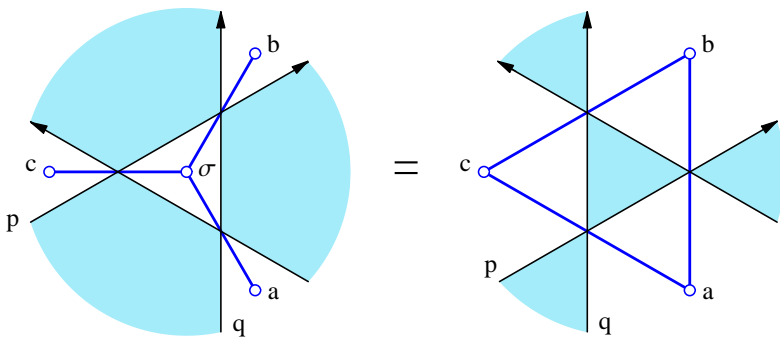}
\end{center}
\caption{A pictorial representation of the star-triangle relation \eqref{str}.}
\label{startriangle}
\end{figure}

The partition function \eqref{Z-def1}
possesses remarkable invariance properties \cite{Bax1,VJones,Bax2}.
It remains unchanged
by continuously deforming the lines of ${\mathscr L}$
(with their boundary
positions kept fixed) as long as the graph ${\mathscr L}$
remains directed. In
particular, no closed directed path are allowed to appear%
\footnote{Actually, these restrictions can be removed if one properly defines
``reflected'' rapidities for downward going lines (see Sect.3 of
\cite{Bax2}), but we will not elaborate this point here here.}%
.
It is easy to see that all such transformations
reduce to a  combination of the moves shown in Fig.~\ref{fig-i1},
Fig.~\ref{fig-i2} and Fig.~\ref{startriangle}.
The special role of the star-triangle \eqref{str} and inversion \eqref{inv}
relations is that they guarantee the invariance of the partition
function under these moves. Given that the graphs ${\mathscr L}$ and
${\mathscr G}$ can undergo rather drastic
changes, the above ``$Z$-invariance'' statement is rather non-trivial.

The partition function \eqref{Z-def1} depends on the exterior spins
and the rapidity variables $p_1,p_2,\ldots,p_L$. Of course, it also
depends on the graph  $\mathscr L$, but only on a relative ordering
(permutation) of the rapidity lines at the boundaries and not on their
arrangement inside  the graph.
Naturally, this graph can be identified with an element of the braid
group.
Then
the partition function $Z$ can be regarded as the corresponding
representation matrix, acting on the continuous boundary spins.
There are several celebrated appearances of the braid group in the theory of
integrable systems, particularly, in connection with
the Jones polynomials \cite{VJones} and the Tsuchiya-Kanie monodromy
representations in conformal field theory \cite{Tsu}.
In Sect.~5 we will show that the rapidity graph $\mathscr{L}$ also
describes combinatorial properties of integrable circle patterns.

The integral \eqref{Z-def1} is well defined in the {\em physical regime},
when all the rapidity differences $\theta_{(i,j)}$ lie in the
interval $0<\theta_{(i,j)}<\pi$.
It rapidly
converges due to the fast exponential decay \eqref{decay} of the edge
weights.
For other values of $\theta_{(i,j)}$
the partition function \eqref{Z-def1}
is defined with an analytic continuation from the physical regime.
In what follows we will restrict ourselves to the physical regime only
(in this case the graph
$\mathscr L$ cannot contain more than one intersection for the same
pair of lines).

Let the graph $\mathscr{G}$ has a very large number of edges $N$. 
We assume that the number of its exterior sites scales as $\sqrt{N}$ 
and that the boundary spins are kept finite.
Following \cite{Bax1} one can show 
that the  leading asymptotics of the partition function
\eqref{Z-def1} at large $N$ has the form
\begin{equation}
\log Z= -\sum_{(ij)\in E(\mathscr{G})}
f_{edge}(\theta_{(ij)})+O(\sqrt{N})\label{factor}
\end{equation}
where the function $f_{edge}(\theta)$ does not depend on
specific details of the lattice. It will be the same as for
the regular square lattice. This result
holds for any $Z$-invariant system with positive Boltzmann weights for
a large graph $\mathscr L$ in a general position. Then, it follows from
\eqref{fzero}, that in the Faddeev-Volkov model {\em with our normalization
of the weights} \eqref{W} the edge free energy vanishes identically,
\begin{equation}
f^{FV}_{edge}(\theta)\equiv 0.\label{FV-zero}
\end{equation}

Consider some additional combinatorial and geometric structures
associated with the graph $\mathscr L$.
First, if the unshaded faces in the above definition of $\mathscr{G}$
are replaced by the shaded ones, one obtains another
graph $\mathscr{G}^*$, which is dual to $\mathscr{G}$.
Each site of $\mathscr{G}^*$ corresponds to a face of $\mathscr{G}$
and vice versa. Obviously, both graphs $\mathscr{G}$ and $\mathscr{G}^*$
have the same medial graph $\mathscr{L}$. Assign the difference
variables $\theta_{e^*}$ to the edges of $\mathscr{G}^*$ by the same
rule \eqref{difvar}. Note that there is one-to-one correspondence between the
edges of $\mathscr{G}^*$ and $\mathscr{G}$. Moreover, if $e\in
E(\mathscr{G})$ is of the first type then the corresponding edge $e^*\in
E(\mathscr{G}^*)$ is of the second (and vice versa).  In other words for the
corresponding edges $\theta_e+\theta_{e^*}=\pi$.
Let ${star}(i)$ denote the set edges meeting at the site
$i$. One can show that for any interior site of $\mathscr{G}$
\begin{equation}
\sum_{(ij)\in {star}(i)} \theta_{(ij)}=2\pi,\qquad
i\in V_{int}(\mathscr{G})\ .\label{sumrule1}
\end{equation}
The similar sum rule holds for the dual graph ${\mathscr G}^*$,
\begin{equation}
\sum_{(kl)\in {star}(k)} \theta_{(kl)}=2\pi, \qquad
k\in V_{int}(\mathscr{G}^*)\label{sumrule2}
\end{equation}

There is a {\it dual} statistical mechanics model with spins $\sigma^*_i$
attached to the sites
$i\in V(\mathscr{G}^*)$.  Its partition function is defined as
\begin{equation}
Z^*=\int  \prod_{(i,j)\in E({\mathscr G}^*)}
W_{\theta^*_{(i,j)}}(\sigma_i^*-\sigma_j^*)\  \prod_{i\in
V_{int}({\mathscr G}^*)} d\sigma_i^*\;,\label{Z-def2}
\end{equation}
where the spins at the exterior sites of $\mathscr{G}^*$
are fixed. Evidently,
it possesses exactly the same invariance properties under
the  deformations of $\mathscr{L}$, as the partition function
\eqref{Z-def1}. Using the standard arguments \cite{Baxterbook} based
on the duality transformation $\eqref{Wp}$ one can
relate \eqref{Z-def2} with \eqref{Z-def1}.

\begin{figure}[hbt]
\begin{center}
\includegraphics[scale=0.6]{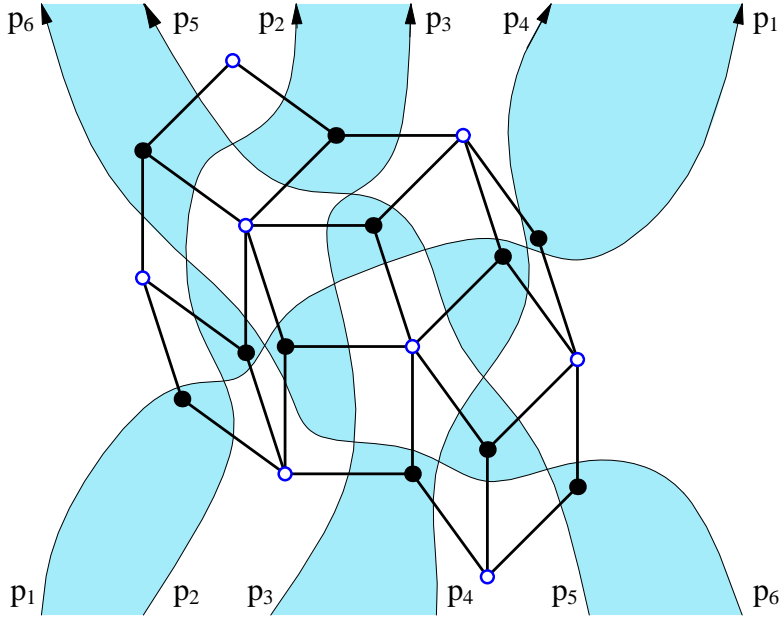}
\end{center}
\caption{A rhombic embedding of the graph ${\mathscr L}^*$}
\label{fig6}
\end{figure}
\begin{figure}[hbt]
\begin{center}
\includegraphics[scale=0.6]{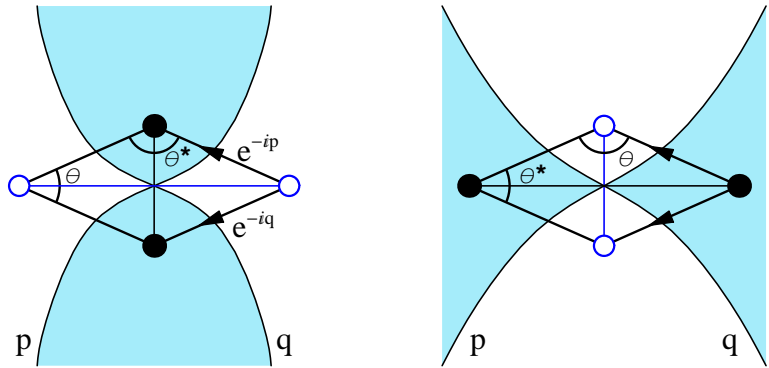}
\end{center}
\caption{Two types of rhombi.}
\label{two-rhombi}
\end{figure}

Consider yet another graph $\mathscr{L}^*$, dual to $\mathscr{L}$.
The set of sites of $\mathscr{L}^*$ consists of those
of $\mathscr{G}$ and $\mathscr{G}^*$.
They are shown in Fig.~\ref{fig6} by white and black dots,
respectively.
The edges of ${\mathscr L}^*$ always connect
one white and one black site.
%
The faces of $\mathscr{L}^*$ corresponds to the vertices of $\mathscr{L}$.
Since the latter are of the degree four,
all faces of $\mathscr{L}^*$ are quadrilateral.
The edges of $\mathscr{G}$ and $\mathscr{G}^*$ are diagonals of these
quadrilateral (see Fig.~\ref{two-rhombi}).
Remarkably, the graph $\mathscr{L}^*$
admits a {\em rhombic embedding} into the plane.
In other words this graph can be drawn so that
all its edges
are line segments of the same length and, hence, all its
faces are rhombi, as shown in Fig.~\ref{fig6}.
The corresponding theorem has been
recently proven in \cite{Kenyon}.
It states that such embedding exists
if and only if (a) no two lines of $\mathscr L$ cross more than
once\footnote{%
The lines of $\mathscr L$ are
called ``train tracks'' (!) in \cite{Kenyon}.}
and  (b) no line of
$\mathscr L$ crosses itself or periodic.
Note, that in the physical regime these conditions are obviously satisfied.

Assume the edges of the quadrilaterals to be of the unit length and consider
them as vectors in the complex
plane.
To precisely specify a rhombic embedding one needs to provide
angles between these vectors.
A rapidity line always crosses opposite (equal)
edges of a rhombus.
Therefore, all edges crossed by same rapidity
line $p$ are equal to each other.
They are determined by
one vector $\zeta_p$,\  $|\zeta_p|=1$.
Thus, if the original rapidity graph ${\mathscr L}$
has $L$ lines, there will be only $L$
different edge vectors.
Choose them as $\zeta_{p_k}=e^{-ip_k}$, where
$p_1,p_2,\ldots,p_L$ are the corresponding rapidity variables.
Each face of $\mathscr{L}^*$ is crossed by exactly two rapidity lines
$p_k$ and $p_\ell$.
To this face associate a rhombus with the edges
$\zeta_{p_k}$ and $\zeta_{p_\ell}$, as shown in Fig.~\ref{two-rhombi}.
Its diagonals are
edges of $\mathscr{G}$ and $\mathscr{G}^*$.
The rhombus angles are
precisely the ``difference variables'' $\theta_e$ and $\theta_{e^*}^*$
assigned to these edges (this is true for both types of rhombi shown
in Fig.~\ref{two-rhombi}).
Glue two rhombi together if they are adjacent.
In the physical regime all their
angles will be in the range from $0$ to $\pi$. So the rhombi will have
a positive area and, thus, will not overlap.
The sum rules \eqref{sumrule1} and \eqref{sumrule2} guarantee that
the resulting surface is flat with no cusps at the sites of ${\mathscr
  L}^*$.

\section{Quasi-classical limit}

Consider the limit $\crs\to\infty$ which corresponds to the
quasi-classical limit of the model.  The parameter $\crs$ becomes
large when $b\to0$ or $b\to\infty$. For definiteness assume $b\to0$,
then $\crs\simeq (2b)^{-1}$.
The weight function $W_\t(s)$
acquires a very narrow bell-shaped form and rapidly decays outside
a small interval $|s|<b/\t$. The integral \eqref{Z-def1} can be then
calculated with a saddle point method. The asymptotic expansion
of the Boltzmann weight \eqref{W} reads
\begin{equation}
W_{\theta}\Big(\frac{\rho}{2\pi b}\Big)
\;=\; \exp\left\{ -\frac{1}{2\pi b^2} A(\theta\/|\rho) +
B(\theta\/|\rho) +
O(b^2)\right\}\;,\qquad b\to 0\ , \label{asyexp}
\end{equation}
where
\begin{equation}
A(\theta\/|\rho)\;=\;A(\theta\/|-\rho)\;=\;\frac{1}{\ii} \int_0^\rho
\log\left(\frac{1+\EXP^{\xi + \ii \theta}}{\EXP^{\xi} +
\EXP^{\ii \theta}}\right)\, d\xi\ .\qquad \label{A-def}
\end{equation}
Note that $A(\theta\/|\rho)\ge0$. It is a smooth  even function of $\rho$
with the minimum value $A(\theta\/|0)=0$ reached at $\rho=0$.
It is simply related
to the Euler dilogarithm function
\begin{equation}
A(\theta\/|\rho)=i \,{Li}_2\left(-\EXP^{\rho-\ii \theta}\right) -i\,
{Li}_2\left(-\EXP^{\rho+\ii \theta}\right) -\theta\/\rho, \qquad
{Li}_2(x)=-\int_0^x\frac{\log(1-x)}{x}dx\ .\label{euler}
\end{equation}
The constant term in \eqref{asyexp} reads
\begin{equation}
B(\theta\/|\rho)=
-\frac{\theta}{2\pi}\,\frac{\partial}{\partial \theta}\,A(\theta\/|\rho)
+\frac{1}{2\pi} \int_0^{\theta} \frac{zdz}{\sin z}\ . \label{bterm}
\end{equation}
Estimating the integral \eqref{Z-def1} for small $b$  one gets
\begin{equation}
\log Z= -\frac{1}{2\pi b^2}\,{\mathscr
  A}[\rho^{(cl)}]\,+\, {\mathscr B}[\rho^{(cl)}]+O(b^2)\ ,\label{zas}
\end{equation}
where
\begin{equation}
{\mathscr A}[\rho\,]=\sum_{(ij)\in E({\mathscr G})}
A\left(\theta_{(ij)}\,|\,\rho_i-\rho_j\right) \label{cl-action}
\end{equation}
and
\begin{equation}
{\mathscr B}[\rho\,]=-\frac{1}{2}\log\det\Big\Vert\frac{\partial^2{\mathscr
    A}[\rho\,]}{\partial\rho_i\partial\rho_j}\Big\Vert \  +
\sum_{(ij)\in E({\mathscr G})}
B\left(\theta_{(ij)}\,|\,\rho_i-\rho_j\right). \label{det}
\end{equation}
Here we use rescaled spin variables $\rho=\{\rho_1,\rho_2,\ldots\}$, \
given by $\rho_i=2\pi b\sigma_i$. The symbol
$\rho^{(cl)}$ denotes the stationary point of the action \eqref{cl-action},
defined by the classical equations of motion
\begin{equation}
\frac{\partial {\mathscr A}[\rho\,]}{\partial
  \rho_i}\Big\vert_{\rho=\rho^{(cl)}} =0,\qquad i=\in
  V_{int}({\mathscr G})\ .
\label{hirota1}
\end{equation}
Explicitly, they read
\begin{equation}
\prod_{(ij)\in star(i)} \;
\frac{\EXP^{\rho_j}+\EXP^{\rho_i+\ii\theta_{(ij)}}}
{\EXP^{\rho_i}+\EXP^{\rho_j+\ii\theta_{(ij)}}}\;=\;1\;,
\qquad i=\in V_{int}({\mathscr G})\ ,
\label{hirota2}
\end{equation}
where the product is taken over the star of edges connected to the
site $i$.
These are the so-called cross ratio equations \cite{BoSur}.
They arise in connection with the circle
patterns \cite{BoSur}, which will be considered in the next section.

The cross ratio equations \eqref{hirota2} are
closely related to the Hirota difference equation \cite{hirota3}.
As an illustration consider the case regular square lattice ${\mathscr
  G}$ shown in Fig.~\ref{fig0}.
\begin{figure}[hbt]
\begin{center}
\includegraphics[scale=.6]{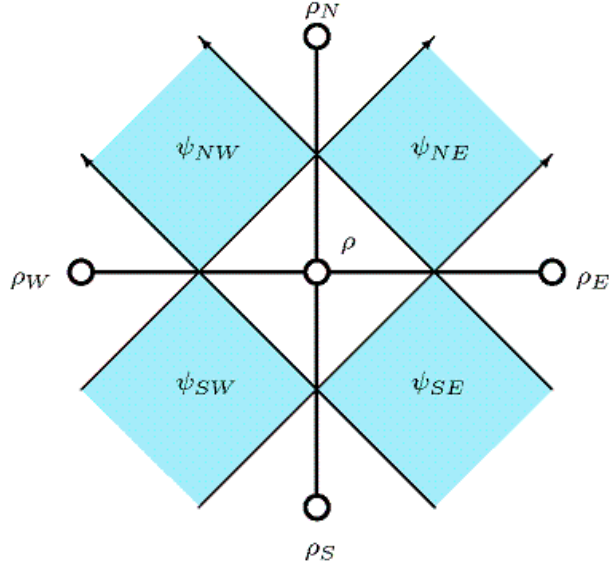}
\end{center}
\caption{A star four edges on the square lattice. The $\psi$-variables
are placed in the sites of the dual lattice, located in the shaded areas.}
\label{forsters}
\end{figure}
A typical ``star'' of four edges is shown in
Fig.~\ref{forsters}. The corresponding equation in \eqref{hirota2} takes
the form
\begin{equation}
\left(
\frac{\EXP^{\rho_W}+\EXP^{\rho+\ii\theta}}{\EXP^{\rho}
+\EXP^{\rho_W+\ii\theta}}
\right)
\left(
\frac{\EXP^{\rho_N}+\EXP^{\rho+\ii\theta^*}}{\EXP^{\rho}
+\EXP^{\rho_N+\ii\theta^*}}
\right)
\left(\frac{\EXP^{\rho_E}+\EXP^{\rho+\ii\theta}}{\EXP^{\rho}
+\EXP^{\rho_E+\ii\theta}}
\right)
\left(\frac{\EXP^{\rho_S}+\EXP^{\rho+\ii\theta^*}}{\EXP^{\rho}
+\EXP^{\rho_S+\ii\theta^*}}\right)
\;=\;1\;,\label{hirota3}
\end{equation}
where $\theta=p-q$ and $\theta^*=\pi-\theta$. Now place a purely
imaginary variable $\psi_i$, \ $0\le {\rm Im}\,\psi_i<2\pi$, on every
site $i$ of the dual lattice.
These sites are shown
by black dots located in shaded areas in Figs.~\ref{forsters} and
\ref{two-faces}. Connect these new variables to the existing
variables $\rho_i$ by the following relations.
\begin{figure}[hbt]
\begin{center}
\includegraphics[scale=.5]{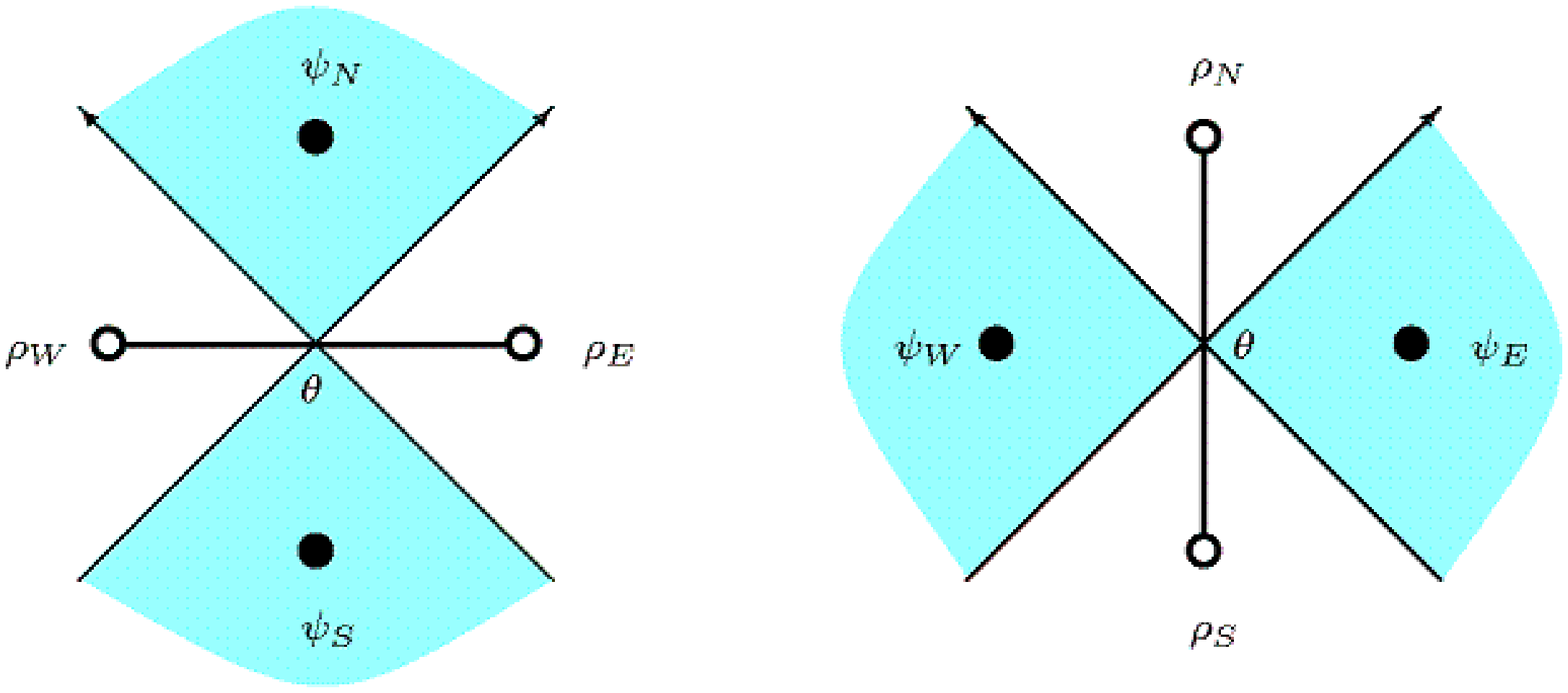}
\end{center}
\caption{Faces of the graph ${\mathscr G}^*$ associated with the
  equations \eqref{hirota4} and \eqref{hirota5}.}
\label{two-faces}
\end{figure}
Let $\psi_N$ and
$\psi_S$ be the variables located above and below of a horizontal edge,
as in Fig.~\ref{two-faces}. Then, we require that
\begin{equation}
\EXP^{\psi_N-\psi_S}=
\frac{\EXP^{\rho_W}+\EXP^{\rho_E+\ii\theta}}{\EXP^{\rho_E}
+\EXP^{\rho_W+\ii\theta}}\ .\label{hirota4}
\end{equation}
Similarly for a vertical edge,
\begin{equation}
\EXP^{\psi_E-\psi_W}=\frac{\EXP^{\rho_N}+\EXP^{\rho_S+\ii\theta^*}}{\EXP^{\rho_S}+\EXP^{\rho_N+\ii\theta^*}}
\quad \Rightarrow \quad\EXP^{\rho_N-\rho_S}=
\frac{\EXP^{\psi_W}+\EXP^{\psi_E+\ii\theta}}{\EXP^{\psi_E}+\EXP^{\psi_W+\ii\theta}}\;.\label{hirota5}
\end{equation}
The consistency of these definitions across the lattice is provided by
\eqref{hirota3}.
Note that the second form of \eqref{hirota5} is identical to
\eqref{hirota4} upon interchanging all $\psi$- and $\rho$-variables.
Therefore it is natural to associate this universal
equation with every face of the
graph ${\mathscr   L}^*$ (there are two type of faces, as shown in
Fig.~\ref{two-rhombi}).

This the famous Hirota equation \cite{hirota3}, which was the starting point
of various considerations in \cite{FV:1994}. A special feature of the
present case
is that the $\rho$-variables are real $\rho\in {\mathbb R}$, while
the $\psi$-variables are purely imaginary, $|\EXP^{\psi_i}|=1$.

\subsection{Star-triangle relation}
Before concluding this section let us discuss some consequences of the
$Z$-invariance. Obviously, this property
applies to every term of the quasi-classical expansion \eqref{zas}.
As an example, consider the star-triangle relation \eqref{str}.
Introduce new variables
\begin{equation}
\t_1=\pi-q+r,\qquad \t_2=\pi-p+q,\qquad \t_3=p-r
\end{equation}
such that
\begin{equation}\label{2pi}
\theta_1+\theta_2+\theta_3\;=\;2\pi\;.
\end{equation}
  Substituting
\eqref{asyexp} into \eqref{str} and using the symmetry
\eqref{crossing-sym} one obtains
\begin{equation}\label{str-semi}
\int \frac{d\rho_0}{2\pi b} \exp\left\{ -\frac{1}{2\pi b^2}
\mathscr{A}_{{\bigstar}}[\rho] + \mathscr{B}_{\bigstar}[\rho]\right\}\;=\;
\exp\left\{-\frac{1}{2\pi b^2} \mathscr{A}_{\triangle}[\rho] +
\mathscr{B}_{\triangle}[\rho]+O(b^2)\right\}
\end{equation}
where
\begin{equation}
\mathscr{A}_{\bigstar}[\rho]\;=\;A(\theta_1|\rho_0-\rho_1)+
A(\theta_2|\rho_0-\rho_2)+A(\theta_3|\rho_0-\rho_3)\;,\label{A-star}
\end{equation}
\begin{equation}
\mathscr{A}_{\triangle}[\rho]\;=\;A(\pi-\theta_1|\rho_2-\rho_3)+
A(\pi-\theta_2|\rho_3-\rho_1)+A(\pi-\theta_3|\rho_1-\rho_2)\;.\label{A-tri}
\end{equation}
Expressions for $\mathscr{B}_\bigstar$ and $\mathscr{B}_\triangle$
are defined in a similar way, with the function $A(\t|\rho)$ replaced by
$B(\t|\rho)$, given by \eqref{bterm}.
The star-triangle relation \eqref{str-semi} implies two non-trivial identities
valid for arbitrary values of $\rho_1,\rho_2,\rho_3$. The first one is
\begin{equation}
\mathscr{A}_\bigstar [\rho_0^{(cl)},\rho_1,\rho_2,\rho_3] \;=\;
\mathscr{A}_\triangle[\rho_1,\rho_2,\rho_3], \label{12term}
\end{equation}
where $\rho_0^{(cl)}$ is the stationary point of the integral in
\eqref{str-semi}
\begin{equation}
\rho_0^{(cl)}=\log\left(\frac{r_2\,r_3\,\sin \t_1+r_1\,r_3\,\sin \t_2
  +r_1\,r_2\,\sin \t_3}
{r_1\,\sin \t_1+r_2\,\sin \t_2 +r_3\,\sin \t_3}\right),\qquad
  r_i=\EXP^{\rho_i}\ .\label{spoint}
\end{equation}
With an account of \eqref{euler} it is a ``twelve-term'' dilogarithm
identity, which can be easily verified by using the definition of the
dilogarithm. A geometric meaning of this identity in terms of volumes
of certain polyhedra in the Lobachevskii 3-space is explained in
the Appendix~A.
The second identity
\begin{equation}
\frac{1}{2}\,\log\frac{\partial^2 \mathscr{A}_{\bigstar}[\rho]}{\partial
  \rho_0^2}\Big\vert_{\rho_0=\rho_0^{(cl)}}
\;=\;\mathscr{B}_\bigstar[\rho]_{\rho_0=\rho_0^{(cl)}} -
\mathscr{B}_\triangle[\rho]
\end{equation}
is an algebraic identity for $r_1,r_2,r_3$, which follows from
elementary (but lengthy) calculations with the use of
the explicit expression \eqref{spoint} and the relation
\begin{equation}
\int_0^\theta \frac{\xi d\xi}{\sin\xi} - \int_0^{\pi-\theta}
\frac{\xi d\xi}{\sin\xi} \;=\; \pi\log\tan\frac{\theta}{2}\;.
\end{equation}

Note that exactly the same quasi-classical star-triangle relation
\eqref{str-semi} arises in a related, but different,
context of the chiral Potts model
\cite{BazCP}.

\section{Circle patterns and discrete conformal transformations}
The Faddeev-Volkov model on a Z-invariant lattice has a remarkable
connection with the circle patterns related to the discrete
Riemann mapping theorem \cite{T1, RS, HS, S}.
The classical Riemann mapping theorem
states that for any simply connected open domain in the
complex plane (which is not the whole plane) there exists a
biholomorphic mapping from that domain to the open unit disk.
The fact  that the mapping is biholomorphic implies that it is a
conformal transformation.
Likewise, the discrete
Riemann mapping theorem involves  discrete analogs of
conformal transformations. Below we give a short introduction
into this subject\footnote{%
We are indebted to A.~Bobenko and B.~Springborn for many illuminating
explanations, some of which are used below.}.

In the continuous case the conformal
transformations possess two key properties: (i) they
preserve the angles  and (ii) uniformly rescale all infinitesimal lengths
in a vicinity of every point, where the scale depends on the
point.
Now one needs to reformulate these properties for a discrete case, when the
continuous plane is replaced, for example, by a polygonal lattice.
For the first property one could, naively, suggest that the angles
between the edges meeting at the same site remain unchanged.
We will call this the rule (i).
The second property is more complicated, since there is no a lattice
analog of the
infinitesimal length.
Nevertheless, as a first attempt it is
reasonable to try the simplest possibility and suggest that the
lengths of the edges attached to the same site are scaled equally
(though the scale can vary from site to site). We will call this the rule (ii).

Now consider transformations of the ordinary square lattice,
implementing both rules at every site.
The rule (i) requires that the $90^\circ$ angles between the four edges
at any site remain unchanged.
The rule (ii) requires that these edges also
remain equal in length for every site
(though, in principle, the lengths could change from site to site).
A very little inspection shows these
requirements only allow a uniform dilatation
of the whole lattice (of course, the same would be true for any polygonal
lattice, not just the square one).
Obviously, the rules are
too restrictive and do not allow any interesting transformations of
the lattice; they need to be relaxed.
\begin{figure}[hbt]
\begin{center}
\includegraphics[scale=0.51]{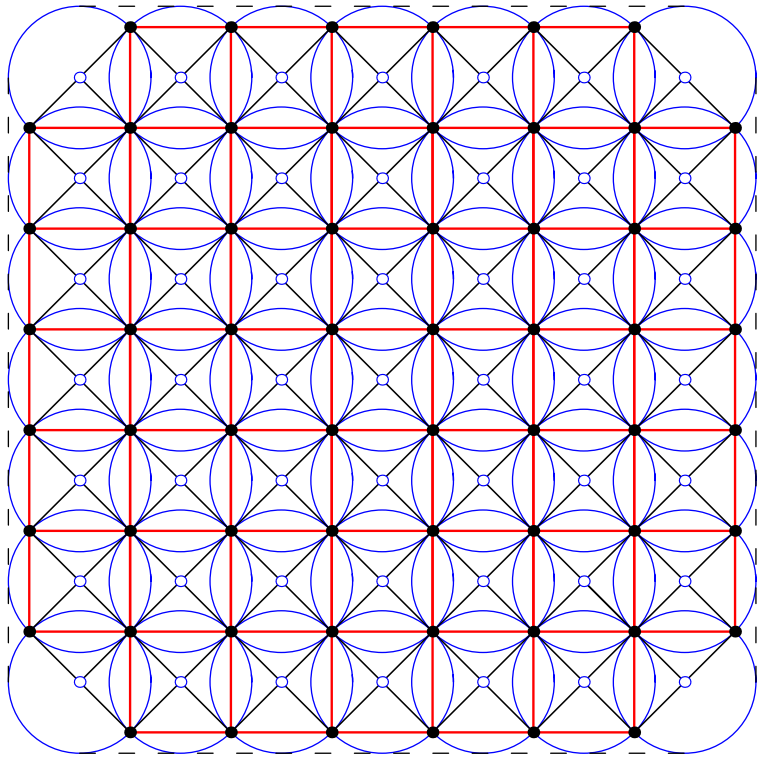}
\end{center}
\caption{The regular square lattice and an associated circle pattern.}
\label{square-lattice}
\end{figure}

A fruitful idea is to work with a bipartite lattice and apply only one
rule on each sublattice.
Consider, for instance, the square lattice shown in
Fig.~\ref{square-lattice}. Its sites are divided into
two sublattices shown with ``white'' and ``black'' dots. Denote them
as ${\mathscr G}$ and ${\mathscr G}^*$ respectively (to unload the
picture the edges of ${\mathscr G}$, connecting the white sites,
are not shown).
There are circles centered at every white site and
passing through four neighboring black sites.
Thus every face of black lattice, ${\mathscr G}^*$, is inscribed in a circle.
Two circles intersect at two distinct points (not just touching)
only if the corresponding faces are adjacent. In the case of
Fig.~\ref{square-lattice} they
always intersect at the right angle, which is the angle between the
edges normal to the circles.

\begin{figure}[hbt]
\begin{center}
\includegraphics[scale=0.55]{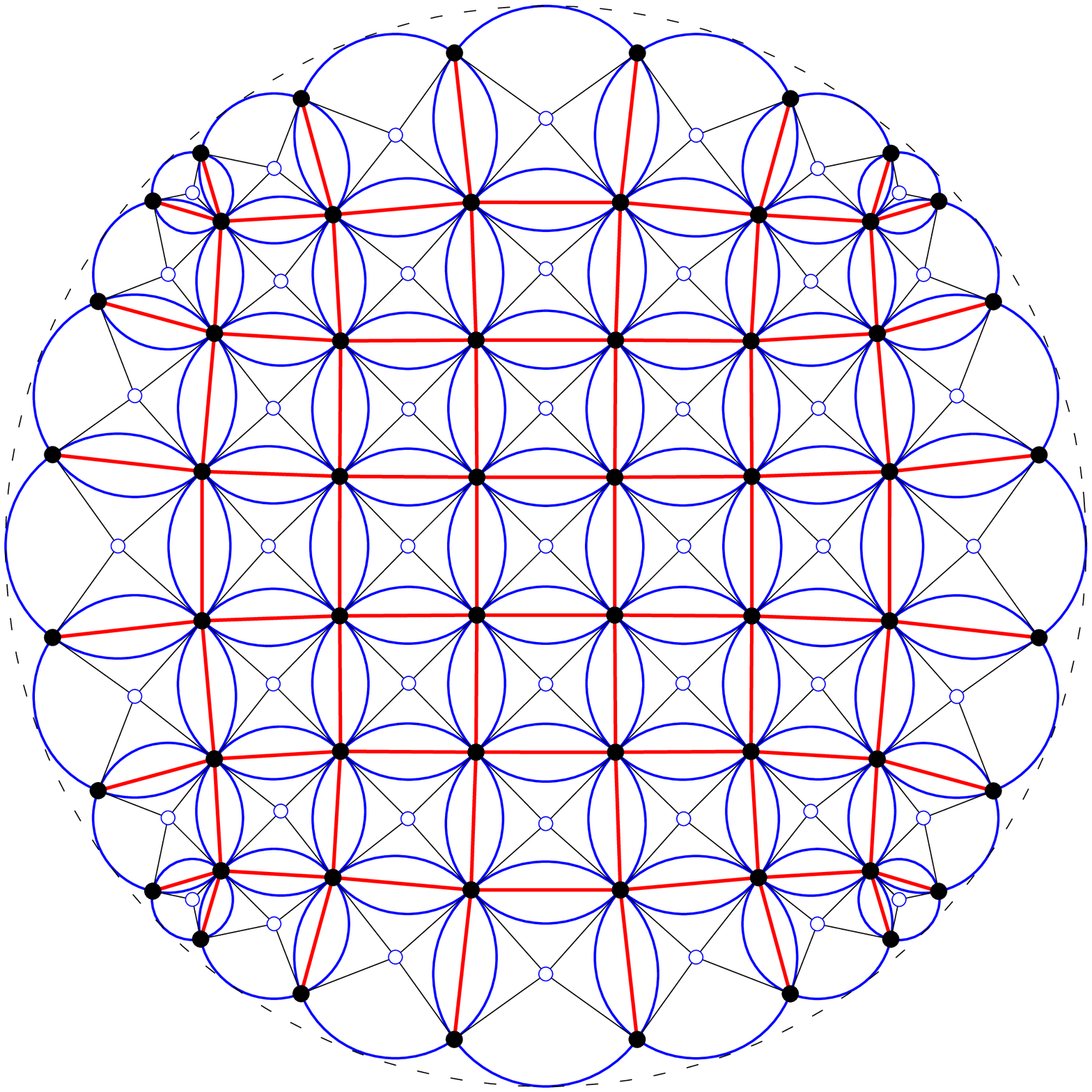}
\end{center}
\caption{A circle pattern with the combinatorics of the square lattice.}
\label{square-circle}
\end{figure}
Let us now distort our lattice implementing the rule (i) for all black
and rule (ii) for all white sites. A particular example of such
transformation is shown in Fig.~\ref{square-circle}. As seen from the
picture,
the circle centered at white sites still pass through all neighboring
black sites (rule (ii)), they still intersect at the $90^\circ$ angles
(rule (i)), but the radii of these circles have changed!
It is important to keep in mind that, by construction, this
transformation preserves the combinatorics of the original lattice. In
particular, there is a one-to-one correspondence between the
circles on the two pictures, which preserves their adjacency
and intersection angles.

Before formulating equations determining the radii of the
circles let us make a few remarks.
Denote by  ${\bf z}=\{z_1,z_2,\ldots,z_N\}$
positions of the circle centers in
the complex plane on the original square lattice, where $N$ is the
total number of circles.
In the case of
Fig.~\ref{square-lattice} this number is $49$.
The circles there cover
a square-like region, bounded by a dashed line.
Let $\{z'_1,z'_2,\ldots,z'_N\}$ be the centers of the {\it corresponding}
circles in Fig.~\ref{square-circle}.
The transformed circle
pattern covers a disk-like region, bounded
by a large circle, also shown by the dashed line%
\footnote{The pattern of
  Fig.~\ref{square-circle} is uniquely fixed (up to a uniform rotation)
by the four-fold axial symmetry
and a requirement that all exterior circles
there touch one additional circle.}%
.
Now define a function $f$ such
that $z'_k=f(z_k)$, for all $z_k\in {\bf z}$. This is a discrete
approximation to the continuous Christophel-Schwartz map.  When $N$
tends to infinity, $f(z)$ uniformly converges to the latter with
an accuracy $O(N^{-1})$ \cite{S,Mat,HS}.
\begin{figure}[hbt]
\begin{center}
\includegraphics[scale=0.55]{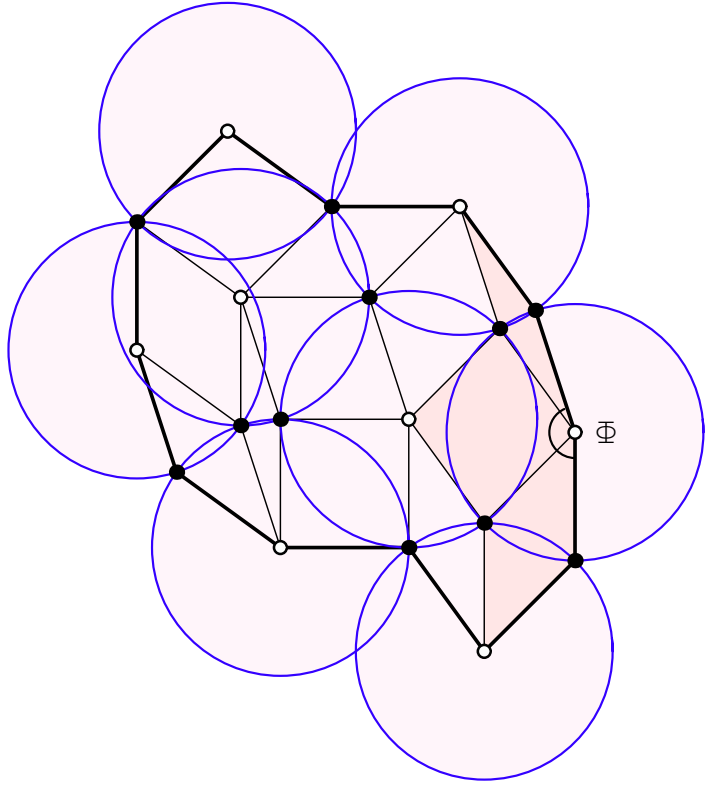}
\end{center}
\caption{An isoradial circle pattern obtained from the rhombic tiling
  of Fig.~\ref{fig6}. The symbol $\Phi$ denotes the cone angle at an
  exterior site.}
\label{pattern1}
\end{figure}
\begin{figure}[hbt]
\begin{center}
\includegraphics[scale=0.6]{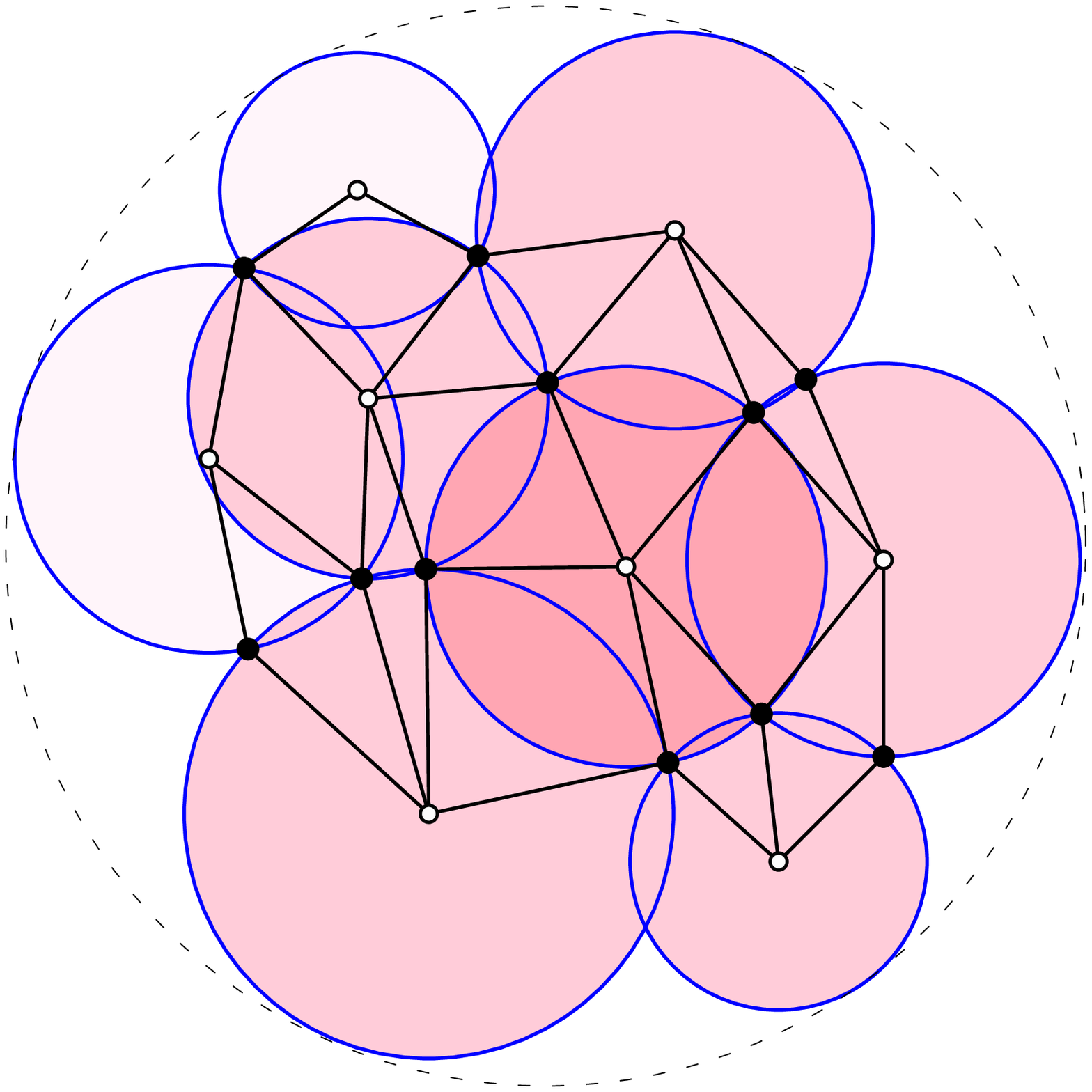}
\end{center}
\caption{An integrable circle pattern with combinatorics of the
graph $\mathscr G$ from Fig.~\ref{fig-net2}. The sum of the kite
angles at the center of the circle flower is equal to $2\pi$.}
\label{pattern2}
\end{figure}

\begin{figure}[hbt]
\begin{center}
\includegraphics[scale=0.4]{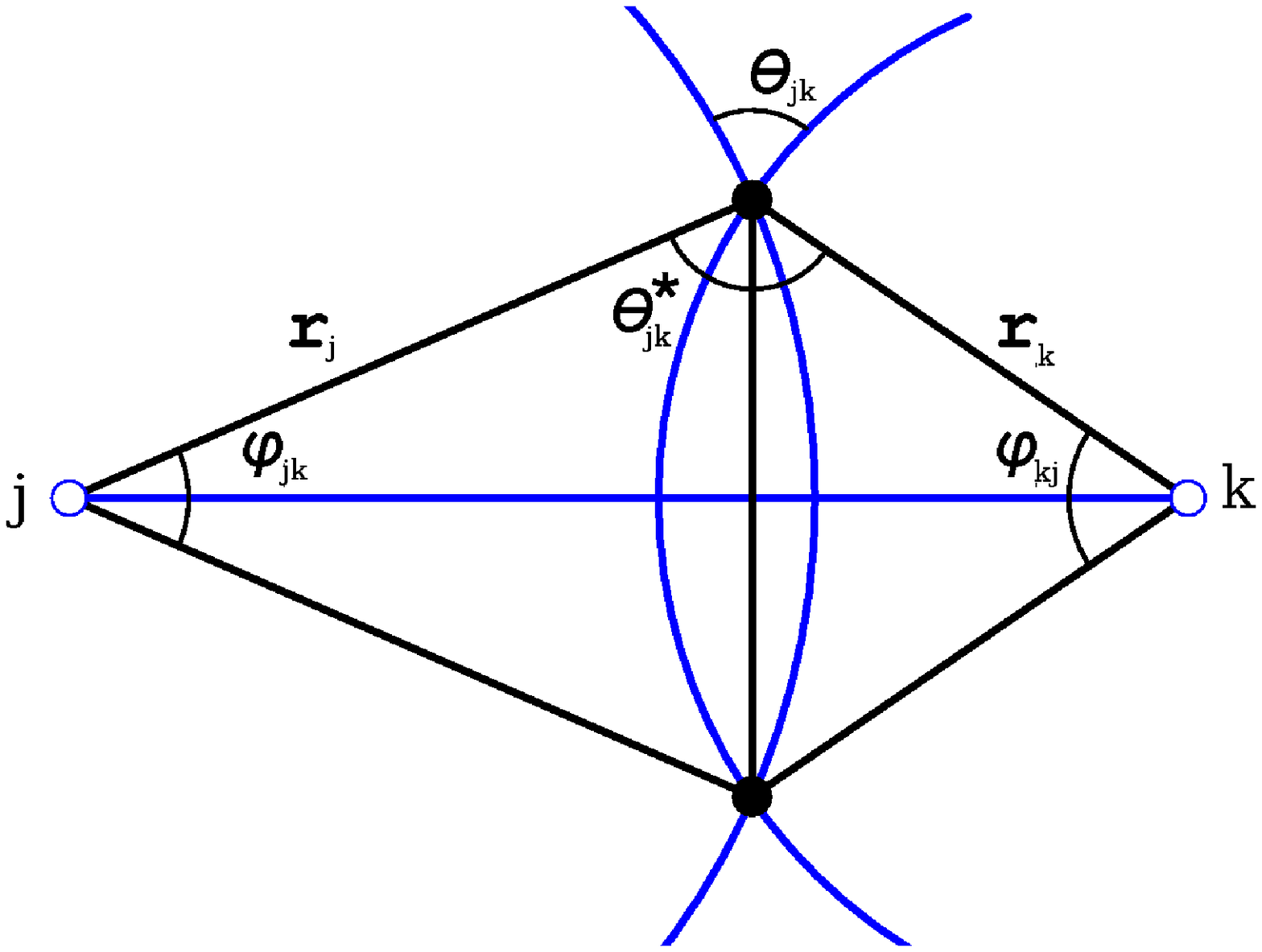}
\end{center}
\caption{Two intersecting circles with the centers at the sites $j$
  and $k$. The angle $\theta_{jk}$ is assigned to
the edge connecting these sites.}
\label{twocircle}
\end{figure}

The patterns considered above enjoy the combinatorics the square
lattice. All circles there intersect at the $90^\circ$ angles.
In a general case, one can consider circle patterns with
the combinatorics of an arbitrary  planar graph
$\mathscr{G}$, where the edges are assigned with arbitrary
intersection angles $\{\theta_e\}$, $e\in E(\mathscr{G})$,\ $0<\theta_e<\pi$.
The centers of the circles correspond to the sites of $\mathscr{G}$
(white sites), such that every face of the dual graph $\mathscr{G}^*$
is inscribed in a circle.
Two circles connected by an edge $e\in E(\mathscr{G})$ intersect at the
angle $\theta_e$ assigned to this edge (Fig.~\ref{twocircle}).
Their intersection points are sites of $\mathscr{G}^*$ (black
sites).
Examples of two different circle patterns associated with same graph
$\mathscr{G}$ as in Fig.~\ref{fig-net2} and having the same
intersection angles are shown in
Figs.~\ref{pattern2} and \ref{pattern1}.
Thus, to define
the {intersection properties} of a  circle pattern (its
combinatorics and intersection angles) one needs to specify a set
$\{\mathscr{G},\{\theta_e\}\}$.

The intersection properties are called {\em integrable} \cite{BMS:2005}
if and only if the corresponding circle pattern
admits an isoradial realization (where all its circles are
the same). Here we will only consider this case.
The integrable circle patterns were extensively studied in connection
with various approximation problems for the Riemann mappings
which are typically
based on a regular lattice combinatorics \cite{S}.
They also connected with integrable discrete non-linear equations
\cite{BMS:2005}.

The space of all ``integrable sets'' $\{\mathscr{G},\{\theta_e\}\}$
was completely described in \cite{BMS:2005}.
Essentially the same arguments  were already presented
in Section~3 throughout the discussion of the Z-invariant systems.
Remind that our starting point there was a rapidity graph ${\mathscr L}$
(Fig.~\ref{fig-net}) with a set of rapidity variables,
one per each line of ${\mathscr L}$.
Using these data we constructed a rhombic ``tiling'' of the dual graph
${\mathscr L}^*$ (Fig.~\ref{fig6}). From this tiling one
can immediately construct an isoradial circle pattern by drawing
circles of the same radius
centered at the white and passing though the neighboring black sites
(Fig.~\ref{pattern1}).
Conversely, any isoradial circle pattern defines
a certain rhombic tiling from which one can easily reconstruct
\cite{Kenyon} the corresponding rapidity graph\footnote{%
Note, that in general this graph can have a multiply
connected boundary.} together with all rapidity values (up to an
insignificant overall shift). Thus the intersection properties
of any integrable circle pattern can be completely described in terms
of some rapidity graph, arising in the $Z$-invariant models (and vice versa).

Let us now derive the equations determining the radii of the circles.
We assume the same notations as in Section~3. We consider the circle
pattern with the combinatorics of ${\mathscr G}$  and identify the rapidity
difference variables $\theta_{(ij)}$, $(ij)\in E({\mathscr G})$ with
the circle intersection angles.  Remind that these variables obey the
sum rules \eqref{sumrule1} and \eqref{sumrule2}.
Every pair of intersecting circles with the centers $i$ and $j$, \
$i,j\in {\mathscr G}$, and the radii $r_i$ and $r_j$ is associated with
a kite-shaped quadrilateral shown in
Fig.~\ref{twocircle}.  The angles
$\varphi_{jk}$ and $\varphi_{jk}$ bisected by the edge $(jk)$ are
given by  (in the notations defined on the figure)
\begin{equation}
\varphi_{jk}=\frac{1}{i}\log\frac {r_j+r_k e^{i\theta_{jk}}}{r_j+r_k
  e^{-i\theta_{jk}}},\qquad\label{phi-def}
\end{equation}
and $\varphi_{kj}$ is obtained by permuting $r_j$ and $r_k$.
Consider a {\em circle flower} consisting of one central circle and
a number adjacent circles (petals), like the one shaded in
Fig.~\ref{pattern2}. Obviously the kite angles at the center of the
flower add up to $2\pi$
\begin{equation}\label{phi-star}
\sum_{(ij)\in {star}(i)} \varphi_{(ij)}=2\pi,\qquad i\in
V_{int}(\mathscr{G})\ .
\end{equation}
Similarly, a sum of the kite angles at any interior black
site (these are $\theta^*$ angles as in Fig~.\ref{twocircle}) should
be equal to $2\pi$ as well, but this condition
is automatically satisfied in virtue
of \eqref{sumrule2}.  Introduce logarithmic radii $\rho_i=\log r_i$ of
the circles. Substituting \eqref{phi-def} into
\eqref{phi-star} and taking into account the sum rule
\eqref{sumrule1} one precisely obtains the equation \eqref{hirota2}
which arose in the quasi-classical limit to the Faddeev-Volkov model.

The quasi-classical action ${\mathscr A}[\rho]$ in \eqref{cl-action}
correspond to the {\em fixed boundary conditions}, where the radii of
all boundary circles are fixed. They can be chosen arbitrarily.
 Bobenko and Springborn
\cite{BSp} considered {\em free boundary conditions}.
Their action
\begin{equation}
{\mathscr A}_{BS}[\rho]={\mathscr A}[\rho]-\sum_{i\in
  V_{ext}({\mathscr G})} \rho_i \Phi_i +const, \label{Abs}
\end{equation}
involve cone angles $\Phi_i$ at the centers of
exterior circles (see Fig.~\ref{pattern1}).
Their values are arbitrary modulo a single constraints
\begin{equation}
\sum_{i\in
  V_{ext}({\mathscr G})} \Phi_i=-2\pi N_{int} +
2\sum_{e\in E({\mathscr G})} \theta_e\
\end{equation}
where $N_{int}$ is the number of interior sites of ${\mathscr G}$.
In the variation principle for the action \eqref{Abs} one considers the
angles $\Phi_i$ as fixed and the external radii as independent variables.

\section{Conclusion}
In this paper we displayed some remarkable connections of the theory
of integrable quantum systems to discrete geometry. We have shown that
the Faddeev-Volkov model provides  a quantization of the circle
patterns and discrete conformal transformations associated with the
discrete Riemann mapping theorem.

The partition function of the model in the thermodynamic limit is
calculated here via the inversion relation method. It would be
interesting to verify this result by other means (particularly, by
numerical calculation) since the inversion relation method is based on
unproved analyticity assumptions of the ``minimal solution''
\eqref{F}. At the moment not much else is known about the
Faddeev-Volkov model. In particular, the problem of the
diagonalization of the transfer matrix is not yet solved, though some
important advances in the related problems in the continuum
sinh-Gordon were made in \cite{Lukyanov:2001,AlZamolodchikov:2006}.

Finally mention the intriguing relation of the Faddeev-Volkov model
with the hyperbolic geometry discussed in the Appendix~A. The solution
\eqref{W} is constructed from the non-compact quantum dilogarithm
\eqref{fi-def}. Note, that there exist another solution of
Yang-Baxter equation connected to this dilogarithm
\cite{Kashaev:tech}. The latter is related to the link invariants
\cite{Hikami:2006} which are also connected with hyperbolic geometry,
particularly, to the volumes of hyperbolic 3-manifolds
\cite{Kashaev:1997lmp,Mur2}. It would be interesting to understand
these connections further.

\vspace{1cm}

\noindent\textbf{Acknowledgements} The authors would like to thank
L.D.Faddeev, A.Yu.Volkov,
M.T.Batchelor, M.Bortz and X.-W.Guan, R.M.Kashaev, E.K.Sklyanin and
A.B.Zamolodchikov for interesting discussions. Special thanks to
A.Bobenko and B.Springborn for introducing us to a beautiful world
of circle patterns.

\app{Star-triangle relation and hyperbolic geometry}

\begin{figure}[hbt]
\begin{center}
\includegraphics[scale=0.6]{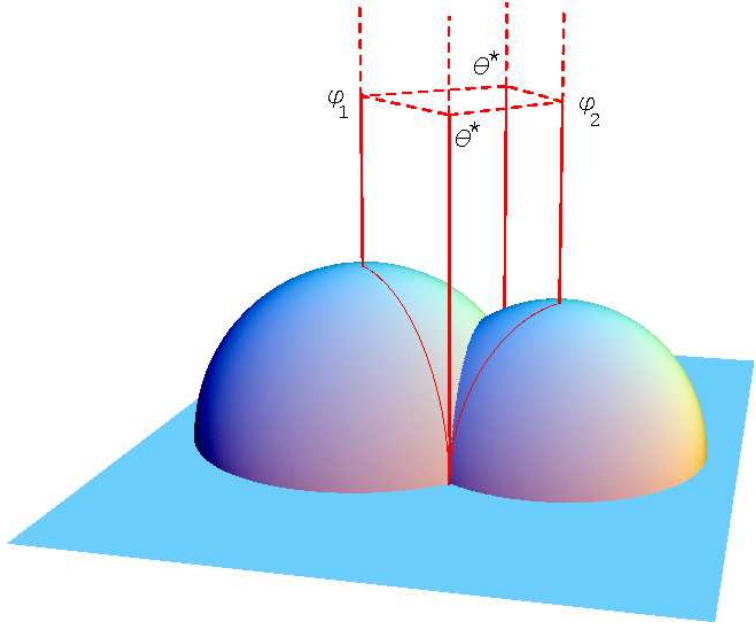}
\end{center}
\caption{A hexahedron in the hyperbolic 3-space bounded by four
  infinite vertical planes and two hemispheres.}
\label{prism}
\end{figure}

An excellent
introduction into the volume calculation in hyperbolic geometry
can be found in
\cite{milnor,vinberg}.
As shown in \cite{springborn, springborn2} the action \eqref{cl-action} has a
geometric interpretation as the volume
of a certain polyhedron in the Lobachevskii 3-space.
Consider the Poincar\'e half-space model $\{x,y,z\in{\mathbb R}|z>0\}$
of the hyperbolic $3$-space with the metric $ds^2=(dx^2+dy^2+dz^2)/z^2$.
In this model hyperbolic planes are represented by hemispheres and
half-planes which intersect the infinite boundary
orthogonally in
circles and straight lines. Take a circle pattern (for instance,
the one shown in Fig.~\ref{pattern1})  and imagine it as lying
in the infinite boundary (the $xy$-plane).
If we erect hemispheres over the circles and orthogonal half-planes over
the prolonged boundary edges (these edges are shown by heavy solid lines
in Fig.~\ref{pattern1}), we obtain a set of hyperbolic planes which
bound a polyhedron. Most of its vertices are at infinity; these
include the circle intersection points (in the $xy$-plane) and one
extra point at the infinite boundary ---
the intersection point of the planes raised
from the boundary edges. Obviously, this polyhedron can be glued from
similar shaped hyperbolic hexahedrons, shown in
Fig.~\ref{prism}, which are associated with
every pair of intersecting circles. Every such hexahedron can be
split into two tetrahedra with three vertices at infinity.

In what follows we will refer to Fig.~\ref{twocircle}, which
corresponds to the projection of the hexahedron
to the ideal, but simplify
the notations there taking $j=1$, \ $k=2$ and denoting
$\varphi_{jk},\varphi_{kj},\theta_{jk},\theta^*_{jk}$  by
$\phi_1,\phi_2,\theta,\theta^*$, respectively.  Note that
$\phi_1+\phi_2=\theta$.  From \eqref{phi-def} one has
\begin{equation}
\EXP^{\ii\phi_1}\;=\;\frac{\EXP^{\rho_1}+\EXP^{\rho_2+\ii\theta}}
{\EXP^{\rho_1}+\EXP^{\rho_2-\ii\theta}}\;.
\end{equation}
where $\rho_i=\log r_i$. Introduce
the Milnor's Lobachevskii function
\begin{equation}
\lobachevski(x)\;=\;-\int_0^x \log |2\sin\xi|\, d\xi\ .
\end{equation}
The volume of the hexahedron in Fig.~\ref{prism} is given by \cite{springborn}
\footnote{Note, that our variables
$\phi_1,\phi_2,\theta$ correspond to $2\phi_1,2\phi_2,\theta^*$ in
  \cite{springborn}.}
\begin{equation}
V_{hexahedron}(\phi_1,\phi_2)\;=\;\lobachevski(\frac{\phi_1}{2})
 +\lobachevski(\frac{\phi_2}{2})
 \;=\; 2\lobachevski(\frac{\theta}{2}) + \frac{1}{2}
A(\theta|\rho_1-\rho_2) + \frac{1}{4}
(\phi_1-\phi_2)(\rho_1-\rho_2)\;,
\end{equation}
where the function $A(\theta\/|\rho)$ is defined in \eqref{A-def}. Then, the
quasi-classical action ${\mathscr A}[\rho^{(cl)}]$ in \eqref{zas}
(remind that it corresponds to the fixed (Dirichlet) boundary
conditions) is given by
\begin{equation}
\mathscr{A}[\rho^{(cl)}]\;=\; 2V(P)-2V(P_0) -
\sum_{i\in V_{ext}(\mathscr{G})} \sum_{(ij)\in star(i)}
(2\phi_{ij}-\theta_{ij})\label{A-vol}
\end{equation}
where $V(P)$ is the volume of the hyperbolic polyhedron $P$
whose projection to the ideal is a circle pattern with a given set of
intersection angles $\theta_{ij}$ and boundary radii
$r_i=e^{\rho_i}$. Further, $V(P_0)$ denotes the value of this volume for the
isoradial case (when all circles have the same radius)
\begin{equation}
V(P_0)=\sum_{(ij)\in E({\mathscr G})} \lobachevski(\frac{\theta_{ij}}{2})\;.
\end{equation}
Thus, up to boundary terms the action \eqref{A-vol} coincides  with
an ``excess volume'' of $P$ with respect to $P_0$. The result
\eqref{factor} and \eqref{FV-zero} implies that if the polyhedron $P$
is glued from a very large number  of hexahedrons, $N$, this excess
volume grows only as $\sqrt{N}$. This is not very surprising:
an arbitrary integrable circle pattern is significantly different from
its isoradial counterpart only at the boundary (indeed, compare
Fig.~\ref{pattern1} and Fig.~\ref{pattern2}).

\begin{figure}[htb]
\begin{center}
\includegraphics[scale=0.7]{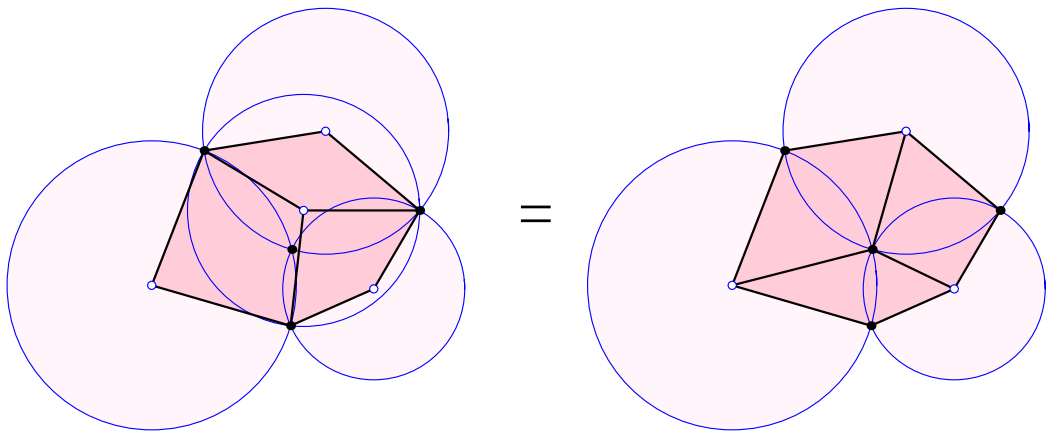}
\caption{Circle patterns corresponding the star (left) and triangular
  sides of the Yang-Baxter equation \eqref{str-semi}.} \label{fig-ybe}
\end{center}
\end{figure}

The star-triangle equation has the following interpretation.
The volume of the polyhedron erected over circle pattern corresponding
to the star side of \eqref{str-semi}, shown in Fig.~\ref{fig-ybe},
is equal to
\begin{equation}
V_{\bigstar}\;=\;2\lobachevski(\frac{\theta_1}{2})+
2\lobachevski(\frac{\theta_2}{2})+2\lobachevski(\frac{\theta_3}{2})\
+ \frac{1}{2}\mathscr{A}_{\bigstar}[\rho_0^{(cl)},\rho_1,\rho_2,\rho_3] +
\textrm{boundary term}\label{V-star}
\end{equation}
where the action $\mathscr{A}_{\bigstar}$ and $\rho_0^{(cl)}$ are
defined in \eqref{A-star} and \eqref{spoint}.
Similarly, the volume of
the polyhedron corresponding to the triangle side of the
star-triangle relation in Fig.~\ref{fig-ybe} is given by
\begin{equation}
V_{\triangle}\;=\;2\lobachevski(\frac{\pi-\theta_1}{2})+
2\lobachevski(\frac{\pi-\theta_2}{2})+2\lobachevski(\frac{\pi-\theta_3}{2})
+ \frac{1}{2}\mathscr{A}_{\triangle}[\rho_1,\rho_2,\rho_3]+
\textrm{boundary term}\label{V-tri}
\end{equation}
where $\mathscr{A}_{\triangle}$ is defined in \eqref{A-tri}.
Clearly, the difference between these two volumes is the volume of an
ideal tetrahedron with its vertices located at the circle
intersection points in Fig.~\ref{fig-ybe}. Indeed, taking into account
\eqref{12term} and the fact that the boundary terms in \eqref{V-star}
and \eqref{V-tri} coincide, one obtains
\begin{equation}
V_{\textrm{tetrahedron}}= V_{\bigstar}-V_{\triangle}=
2\lobachevski(\frac{\theta_1}{2})
+2\lobachevski(\frac{\theta_2}{2})
+\lobachevski(\frac{\theta_3}{2})
-2\lobachevski(\frac{\pi -\theta_1}{2})
-2\lobachevski(\frac{\pi -\theta_2}{2})
-\lobachevski(\frac{\pi -\theta_3}{2})\; .
\label{vtet}
\end{equation}
Using the identity
\begin{equation}
2\lobachevski(\frac{\theta}{2})-2\lobachevski(\frac{\pi-\theta}{2})
=\lobachevski(\theta)\;.
\end{equation}
one obtains
\begin{equation}
V_{\textrm{tetrahedron}}=
\lobachevski(\theta_1)+\lobachevski(\theta_2)+\lobachevski(\theta_3)\;,
\label{vtet1}
\end{equation}
which is precisely the Milnor's formula \cite{milnor}
for the volume of the ideal
tetrahedron. We would like to stress that whereas the volumes
$V_{\bigstar}$ and $V_{\triangle}$ depend on the radii of the circles,
their difference does not. It depends only on the angles $\t_1$,
$\t_2$ and $\t_3$.  In \eqref{12term} this difference is absorbed into
the normalization of the edge weights. Correspondingly, the RHS of
\eqref{vtet} contains six terms, one for each edge of the tetrahedron.
Clearly, the action \eqref{A-vol} is invariant under any star-triangle
transformation since volumes of ideal tetrahedrons will cancel out
from the difference $V(P)-V(P_0)$.

\app{Properties of the functions $\varphi(z)$ and $\Gfun(z)$.}\label{phi-prop}

Below we use the following notations
\begin{equation}
\eta=\frac{1}{2}(b+b^{-1}),\qquad
q=\EXP^{\ii\pi b^2},\qquad \tilde{q}=\EXP^{-\ii\pi b^{-2}},\qquad
\overline{q}=\ii\,\exp\left(\frac{\ii\pi(b-b^{-1})}{2(b+b^{-1})}\right)\ ,
\end{equation}
\begin{equation}
(x,q)_\infty\,\be\, \prod_{k=0}^\infty (1-q^k x)\ .
\end{equation}

\noindent
{\em The function $\varphi(z)$.}
This function is defined by the integral \eqref{fi-def}. It has
the following properties.
\begin{enumerate}[(a)]
\item
{\em Simple poles and zeros}
\begin{equation}\label{poles}
\begin{array}{l}
\textrm{poles of } \varphi(z)\;=\;\left\{
\ii\crs + \ii m b + \ii n b^{-1}\;,\;\;\;m,n\in\mathbb{Z}_{\geq 0}\right\}\;,\\
\textrm{zeros of } \varphi(z)\;=\;\left\{ -(\ii \crs + \ii m b +
\ii n b^{-1})\;,\;\;\;m,n\in\mathbb{Z}_{\geq 0}\right\}\;.
\end{array}
\end{equation}
\item
{\em Functional relations}
\begin{equation}\label{difference}
\varphi(z)\varphi(-z)\;=\;\EXP^{\ii\pi z^2 - \ii\pi
(1-2\crs^2)/6}\;,\qquad  \frac{\varphi(z-\ii b^{\pm 1}/2)}
{\varphi(z+\ii b^{\pm 1}/2)}\;=\; \left(1\,+\,\EXP^{2\pi z
b^{\pm 1}}\right)\;.
\end{equation}
\item
{\em Asymptotics}
\begin{equation}
\varphi(z)\;\simeq\;1,\qquad\Re (z)\to
-\infty;\qquad\varphi(z)\;\simeq\;\EXP^{\ii\pi
z^2-\ii\pi(1-2\crs^2)/6}\qquad \Re (z)\to
+\infty\;.
\end{equation}
where $\Im (z)$ is kept finite.
\item
{\em Product representation}
\begin{equation}
\varphi(z)=\frac{(-q\,e^{2\pi z\,b}\,;\ q^2)_\infty}
{( -\tilde q\,e^{2\pi z\, b^{-1}};\tilde  q^{\,2})_\infty},\qquad \Im\, b^2>0
\end{equation}
\item
{\em Pentagon relation.} The function $\varphi(z)$ satisfy the
following operator identity \cite{Faddeev:1994}
\begin{equation}
\varphi(\mathsf{P}) \varphi(\mathsf{X}) \;=\;
\varphi(\mathsf{X}) \varphi(\mathsf{P}+\mathsf{X})
\varphi(\mathsf{P})\;,\quad
[\mathsf{P},\mathsf{X}]\;=\;\frac{1}{2\pi\ii}\;.\label{pent}
\end{equation}
where $[\ ,\ ]$ denotes the commutator. It can be re-written in the matrix
form \cite{Ponsot:2001,Kas01}
\begin{equation}\label{rstr}
\ds \int_{\mathbb{R}}\; \frac{\varphi(x+u)}{\varphi(x+v)}\;
\EXP^{2\pi\ii wx}\;dx\;=\; \EXP^{\ii\pi(1+4\crs^2)/12\ -2\pi\ii
w(v+\ii\crs)}\
\frac{\varphi(u-v-\ii\crs)\varphi(w+\ii\crs)}{\varphi(u-v+w-\ii\crs)}\;,
\end{equation}
where the wedge of poles of $\varphi(x+u)$ must lie in the upper
half-plane, the wedge of zeros of $\varphi(x+v)$ must lie in the
down half-plane, and the integrand must decay when
$x\to\pm\infty$.
\end{enumerate}

\noindent
{\em The function $\Gfun(z)$.} This function is defined by the integral
\eqref{Fi-def}. It has the following properties.
\begin{enumerate}[(i)]
\item
{\em Simple poles and zeros}
\begin{equation}\label{G-poles}
\begin{array}{l}
\ds\textrm{poles of $\Gfun(z)$} \;=\; \left\{ 2\ii\crs  + \ii m b
+ \ii n b^{-1}\;, \;\;\; m,n\in \mathbb{Z}_{\geq 0}\;,\;\;\;
m+n-|m-n| = 0 \mod 4 \right\}\;,\\
\ds\textrm{zeros of $\Gfun(z)$} \;=\; \left\{ -(2\ii\crs + \ii m b
+ \ii n b^{-1})\;,\;\;\; m,n\in \mathbb{Z}_{\geq 0}\;,\;\;\;
m+n-|m-n| = 0 \mod 4\right\}.
\end{array}
\end{equation}
\item
{\em Functional relations}
\begin{equation}
\Gfun(z)\;\Gfun(-z)\;=\;\EXP^{\ii\pi z^2/2 - \ii \pi
(1-8\crs^2)/12}\;,\quad \Gfun(z+\ii\crs) \Gfun(z-\ii\crs) \;=\;
\varphi(z)\;.
\end{equation}
\item
{\em Asymptotics}
\begin{equation}
\Gfun(z)\simeq 1,\qquad z\to -\infty;\qquad
\Gfun(z)\simeq \EXP^{\ii\pi z^2/2 -
\ii\pi(1-8\crs^2)/12},\qquad z\to +\infty\ .
\end{equation}
where $\Im (z)$ is kept finite.
\item
{\em Product representation}
\begin{equation}
\Phi(z)=\frac{(q^2 e^{2\pi z\,b};q^4)_\infty}
{(\tilde{q}^{\,2}e^{2\pi z\, b^{-1}};\tilde{q}^{\,4})_\infty}
\frac{\ds(-\overline{q} e^{{\pi z}/(2\eta)};
\overline{q}^{\,2})_\infty}
{\ds(\overline{q} \,e^{{\pi z}/(2\eta)};\overline{q}^{\,2})_\infty},
\qquad \Im\, b^2>0\ .
\end{equation}
\end{enumerate}


\def\cprime{$'$} \def\cprime{$'$}

\end{document}